\documentclass[acmsmall]{acmart}

\usepackage[normalem]{ulem}

\usepackage{xcolor}   
\usepackage{colortbl}       

\usepackage{subcaption}
\usepackage{tikz}
\usepackage{pgf-umlcd}
\usetikzlibrary{arrows.meta, positioning, fit, shapes.misc, calc, shadows.blur}
\usepackage{nccmath}
\usepackage{comment}
\usepackage{tcolorbox}
\usepackage{caption}
\usepackage{tcolorbox}
\usepackage{algorithmicx,algpseudocode} 

\definecolor{chgup}{HTML}{E8F5E9}     
\definecolor{chgnew}{HTML}{E3F2FD}    
\definecolor{chgsame}{HTML}{F5F5F5}   
\definecolor{accepted}{HTML}{F3E5F5}  
\definecolor{dropped}{HTML}{FFEBEE} 
\usepackage{graphicx}

\usepackage{listings}
\lstdefinestyle{mystyle}{
    backgroundcolor=\color{white},   
    commentstyle=\color{green},
    keywordstyle=\color{magenta},
    numberstyle=\tiny\color{gray},
    stringstyle=\color{blue},
    basicstyle=\ttfamily\footnotesize,
    breakatwhitespace=false,         
    breaklines=true,                 
    captionpos=b,                    
    keepspaces=true,                 
    numbers=left,                    
    numbersep=5pt,                  
    showspaces=false,                
    showstringspaces=false,
    showtabs=false,                  
    tabsize=2
}

\usepackage[linesnumbered,ruled,vlined]{algorithm2e}
\usepackage{array}			
\usepackage{multirow}

\usetikzlibrary{arrows.meta,positioning,calc}

\usepackage{mathtools}
\usepackage{enumitem}
\usepackage{comment}
\usepackage{amsmath}

\usepackage{placeins}

\usepackage{ifthen}
\usepackage{fontawesome5}
\usepackage{array}
\usepackage{xltabular}
\usepackage{caption}
\usepackage{enumitem}
\usepackage{longtable}
\usepackage{color}
\usepackage{tikz}

\usepackage[utf8]{inputenc}
\usepackage[linesnumbered, ruled, vlined]{algorithm2e}
\usepackage{algorithmicx}

\newboolean{showcomments}
\setboolean{showcomments}{true} 
\ifthenelse{\boolean{showcomments}}
{\newcommand{\nb}[2]{
    \fcolorbox{gray}{yellow}{\bfseries\sffamily\scriptsize#1}
    {$\blacktriangleright$#2$\blacktriangleleft$}
  }
  
}
{\newcommand{\nb}[2]{}
  
}

\newcommand\todo[1]{\nb{TODO}{\textcolor{red}{\textsl{#1}}}}

\newcommand\hs[1]{\nb{Houari}{\textcolor{teal}{\textsl{#1}}}}

\newcommand{\tool}{RAFiC}
\newcommand{\tooll}{RAFiC }
\newcommand{\M}{\ensuremath{\mathcal{M}}}
\usepackage{titlesec}

\titleformat{\section}
  {\normalfont\Large\bfseries}{\thesection}{1em}{\MakeUppercase}

\usepackage{xcolor}
\usepackage{listings}
\usepackage{makecell}
\usepackage{tabularx}
\usepackage{placeins}

\newcommand{\revadd}[1]{#1}

\usepackage{awesomebox}

\usepackage{adjustbox}
\usepackage{framed}
\definecolor{archbg}{HTML}{FFF8E1}  
\definecolor{relbg}{HTML}{E3F2FD}   
\definecolor{attrbg}{HTML}{F3E5F5}  
\definecolor{archframe}{HTML}{FFC107}
\definecolor{relframe}{HTML}{42A5F5}
\definecolor{attrframe}{HTML}{AB47BC}
\definecolor{hdr}{HTML}{37474F}     
\definecolor{fmt}{HTML}{1B5E20}     
\definecolor{blk}{HTML}{004D40}     

\lstdefinelanguage{Prompt}{
  sensitive=true,
  moredelim=**[is][\bfseries\color{hdr}]{<|start_header_id|>}{<|end_header_id|>},
  moredelim=**[is][\bfseries\color{hdr}]{<|end_header_id|>}{<|eot_id|>},
  moredelim=**[is][\bfseries\color{fmt}]{==}{==},           
  moredelim=**[is][\bfseries\color{blk}]{[BEGIN]}{[END]},   
}

\lstdefinestyle{llamaBase}{
  language=Prompt,
  basicstyle=\ttfamily\small,
  breaklines=true,
  columns=fullflexible,
  frame=single,
  framerule=0.6pt,
  showstringspaces=false
}

\lstdefinestyle{llama-arch}{style=llamaBase, backgroundcolor=\color{archbg}, rulecolor=\color{archframe}}
\lstdefinestyle{llama-rel}{ style=llamaBase, backgroundcolor=\color{relbg},  rulecolor=\color{relframe}}
\lstdefinestyle{llama-attr}{style=llamaBase, backgroundcolor=\color{attrbg}, rulecolor=\color{attrframe}}
\usepackage{tabularx,booktabs}
\newcolumntype{Y}{>{\raggedright\arraybackslash}X}

\begin{document}

\title{From Language-Specific to Cross-DSL: Semantic-Role-Guided Model Completion with LLMs}

\author{Meriem Ben Chaaben}

\orcid{0000-0001-8133-0199}
\affiliation{
  \institution{DIRO, Université de Montréal}
  \city{Montréal}
  \country{Canada}
}
\email{meriem.ben.chaaben@umontreal.ca}

\author{Théo Giraudet }
\orcid{0009-0006-8461-2697}
\affiliation{
  \institution{Obeo, Univ Rennes, Inria, CNRS, IRISA	}
  \city{}
  \country{France}
}
\email{theo.giraudet@irisa.fr}

\author{Benoit Combemale}
\orcid{0000-0002-7104-7848}
\affiliation{
  \institution{Univ Rennes, Inria, CNRS, IRISA}
  \city{}
  \country{France}
}
\email{benoit.combemale@irisa.fr}
\author{Houari Sahraoui}
\orcid{0000-0001-6304-9926}
\affiliation{
  \institution{DIRO, Université de Montréal}
  \city{Montréal}
  \country{Canada}
}
\email{sahraouh@iro.umontreal.ca}

\renewcommand{\shortauthors}{Ben Chaaben et al.}

\setcopyright{acmlicensed}
\acmVolume{1} \acmNumber{1} \acmArticle{1} \acmDOI{}

\begin{abstract}

The rapid emergence of large language models (LLMs), trained on vast and diverse corpora, has positioned them as a promising source of domain knowledge to assist modelers. However, leveraging LLMs for model completion remains challenging: suggestions must respect both the syntax of a domain-specific language (DSL) and the intended domain semantics, while remaining consistent with an evolving model.
This paper presents \textbf{\tool}, an LLM-based approach for iterative domain model completion that generalizes across DSLs. It drives completion using metamodel information to ensure syntactic validity and enriches prompts with semantic-role annotations  to better capture the intended modeling role of DSL elements. The approach provides context-aware, incremental guidance during model construction.
We evaluate \tooll through an ablation-based comparative study over three DSLs and two scenarios per DSL to isolate the contribution of syntactic and semantic guidance. Across these six scenarios, semantic role guidance improves contribution rate in every case, and \tooll improves completion effectiveness over weaker prompting configurations in most of them. The largest gains occur for DSLs that are less represented in LLM training data. In these cases, iterative guidance leads to earlier and more stable accepted suggestions, while semantic role guidance improves their relevance.

\end{abstract}



\maketitle
\setlength{\textfloatsep}{6pt}
\setlength{\floatsep}{6pt}
\setlength{\intextsep}{6pt}
\label{sec:intro}

\section{Introduction}

Domain modeling plays a central role in software and systems engineering by providing structured representations of domain concepts \cite{broy2013domain}. In Model-Driven Engineering (MDE), this is achieved through modeling languages, where a metamodel defines the abstract syntax and a concrete syntax provides notations for creating instance models ~\cite{harel2004meaningful,rath2010synchronization,combemale2016engineering,fondement2007concrete}. 
\revadd{
This process, however, is far from straightforward: Domain modeling is a challenging, time-consuming, and error-prone activity. One of its main difficulties lies in defining the boundaries of the domain to be modeled. Modelers must determine which concepts belong to the domain, how they relate to one another, and which relevant concepts or relationships may have been overlooked. Missing or incomplete domain knowledge at this stage can propagate throughout the development process, leading to models that inadequately capture the intended system.
}
\revadd{
This motivates domain model completion as a form of interactive assistance in software engineering processes~\cite{sen2007domain}. Rather than generating an entire model from scratch, completion support helps modelers progressively define and refine the model by recommending candidate concepts, relationships, and refinements that can be inspected, accepted, rejected, or adapted. Such assistance reduces the risk of omitting important domain elements and relationships while leaving the final modeling decisions under the control of the modeler. Importantly, completion is inherently incremental: each modeling decision changes the
state of the model and, consequently, what may be relevant to suggest next.
}

\revadd{
The assistance therefore requires mechanisms that can reason about both the modeling language and the evolving modeling context.} Recent advances in large language models (LLMs), trained on vast corpora, make this perspective increasingly feasible, as they have shown promising capabilities in supporting software engineering tasks, including modeling assistance~\cite{shi2025llms,ben2024toward,chen2023automated}. By providing context-sensitive suggestions during model construction, LLMs offer new opportunities to assist modelers, particularly when the model is incomplete or created by practitioners with different levels of modeling expertise.
\revadd{
However, effective domain model completion cannot rely solely on the generative capabilities of LLMs. Although LLMs have accumulated extensive knowledge about many application domains, they are generally unfamiliar with the domain-specific modeling languages (DSLs) used to represent this knowledge. Recent work has shown that LLMs exhibit weaker performance on low-resource programming languages due to their limited representation in pretraining corpora~\cite{delgado2026much}. The challenge is even greater for DSLs, which are rarely encountered during pretraining. Consequently, although an LLM may understand domain concepts and their relationships, it may not know how these concepts are represented in a particular DSL, nor infer the intended modeling role of its constructs. It may therefore generate suggestions that are plausible at the domain level but incompatible with the structure or intended use of the modeling language.
}

\revadd{
This observation points to a more general challenge: how can an LLM perform model completion when the language used to express the domain may be unfamiliar to it, while the model itself is continuously evolving?
A completion mechanism must communicate to the LLM not only the current state of the partial model, but also the structural and semantic knowledge needed to interpret that model. Moreover, this information cannot be provided only once: as the model changes through accepted suggestions or manual edits, the context used to generate subsequent
suggestions must evolve accordingly.
}

\revadd{
We address this challenge by viewing LLM-based model completion as an
iterative communication process between the modeling environment and the
LLM. At each modeling step, the LLM receives three complementary sources
of information: (i) the structural constraints of the target DSL,
captured by its metamodel; (ii) semantic guidance describing the intended
modeling roles of DSL constructs; and (iii) the current state of the
partial model under construction. The LLM then generates candidate
completions, which are presented to the modeler. Once a suggestion is
accepted or the model is otherwise modified, the resulting model state
becomes part of the context for the next completion step. This closed
loop allows suggestions to adapt to both the language and the evolving
model rather than treating completion as a sequence of independent
generation requests.
}

\revadd{
To provide semantic guidance across heterogeneous DSLs, we use a
lightweight semantic annotation scheme that associates DSL constructs
with generic modeling roles. Such annotations provide explicit cues
about the intended modeling function of constructs beyond their
DSL-specific names, allowing domain knowledge captured by an LLM to be
connected to the abstractions of an unfamiliar modeling language.
Unlike approaches that rely on a fixed prompt or provide guidance only
once ~\cite{chaaben2023towards,Ferrari24}, our approach continuously combines this syntactic and semantic language knowledge
with the dynamically evolving model context.
}


More concretely, we propose \textbf{RAFiC}(Recommender for AI-assisted Framework for domaIn model Completion),
an LLM-based approach for iterative domain model completion across DSLs.
RAFiC realizes this iterative communication process by combining
metamodel-driven structural guidance, semantic-role guidance, and
context-aware orchestration. \revadd{
The metamodel and semantic annotations provide reusable knowledge about
the target DSL, while the evolving model provides the context that is
updated throughout the interaction. This combination enables RAFiC to
generate completion suggestions that are adapted both to the modeling
language and to the current state of the model.
}
\revadd{
RAFiC is designed as an assistance layer integrated into a graphical
modeling environment. The resulting mechanism is intended to generalize across DSLs without
requiring the LLM itself to have been trained on each target language.
Only the DSL-specific metamodel and the mapping between its constructs
and the semantic annotation scheme need to be provided as static
language knowledge. This enables the same completion mechanism to operate
across modeling languages while adapting its recommendations to the
specific structure, semantics, and current state of each model.
}



We evaluate \tooll through an ablation-based comparative study over three DSLs and two scenarios per DSL, selected to differ in their expected presence in LLM pretraining data. Within these six scenarios, semantic role guidance improves contribution rate in every case, and \tooll improves completion effectiveness over weaker prompting configurations in most of them, with the largest gains observed for less familiar DSLs where both iterative guidance and semantic role guidance contribute most.

\paragraph{Contributions.}
\begin{itemize}

    \item We propose \tool, an LLM-based approach for iterative domain model completion that combines metamodel-driven structural constraints with semantic annotations of modeling constructs. 

    \revadd{  \item We introduce a cross-DSL semantic guidance mechanism that maps DSL-specific metamodel elements to generic semantic roles, allowing semantic information to be reused across heterogeneous modeling languages without requiring the LLM to have prior familiarity with each DSL.}

    \item We provide an extensive evaluation through targeted comparisons and ablations over three DSLs and two scenarios per DSL, demonstrating the benefits of iterative guidance over minimal and static prompting, and quantifying the contribution of semantic annotations to completion quality.

    \item We implement \tooll as a proof-of-concept tool integrated into \emph{Sirius Web}, enabling interactive domain model completion within a web-based modeling environment.
\end{itemize}
    
    
    

The remainder of the paper is organized as follows. Section~\ref{sec:background} presents the necessary background. Section~\ref{sec:approach} introduces \tool, detailing its main components and its implementation as a proof of concept. Section~\ref{sec:evaluation} describes the study design, and Section~\ref{sec:results} reports the results. Section~\ref{sec:discussion} discusses the main observations and implications of our approach. Section~\ref{sec:threats} discusses threats to the validity of our work. Section~\ref{sec:related_work} positions \tooll with respect to related work, and Section~\ref{sec:conclusion} concludes the paper.

\section{Background}
\label{sec:background}
\revadd{
This section introduces the core concepts underlying our approach: modeling languages and their structure, semantic annotation in Model-Driven Engineering, domain model completion as a task, and the use of LLMs for generative assistance.}

\paragraph{\textbf{Core Concepts of Modeling Languages.}}
Modeling languages, whether general-purpose or domain-specific, are typically defined through a structured process. This process begins with defining a \textit{metamodel}, which serves as an abstract specification of the modeling language. The metamodel describes the essential \textit{concepts, relationships, and structural constraints} that govern valid model instances~\cite{rath2010synchronization} and acts as a \textit{blueprint} for constructing models within the domain. Once the metamodel is established, a \textit{concrete syntax} provides a representation for expressing models, either \textit{graphical} (e.g., UML diagrams) or \textit{textual} (e.g., DSLs using formal grammar) \cite{fondement2007concrete}, enabling domain experts to create and manipulate models.  

Finally, the modeling framework is enriched with \textit{semantics} and \textit{constraints} to ensure that models are meaningful and valid. \textit{Semantics} define the interpretation and behavior of model elements~\cite{harel2004meaningful}, while \textit{constraints} enforce consistency and correctness, often using languages such as OCL~\cite{richters2002ocl} or embedded validation mechanisms within the DSL.  
By structuring modeling languages around metamodels, concrete syntax, and explicit semantics, software engineers can create robust and reusable models that enhance automation, analysis, and verification in Model-Driven Engineering workflows.

\paragraph{\textbf{Semantic Annotation in Model-Driven Engineering.}}
Semantic annotation is a widely used mechanism in Model-Driven Engineering for making the meaning of model and metamodel elements explicit beyond what their structural definition conveys.

 This idea has been applied across several MDE tasks. In business process modeling (BPMN), ~\citet{lautenbacher2008semantic} annotate process elements such as tasks, roles, and data objects with domain and ontology concepts to make process semantics explicit and support automated validation, while ~\citet{di2015semantic} review and extend methodologies for annotating BPMN process models to support process comparison, retrieval, and reuse across organizations.  In conceptual modeling, the TwoUse approach~\cite{parreiras2010using} annotates UML classes and properties with OWL DL axioms so that object-oriented modeling and description-logic reasoning can operate over the same artifact. Finally, ~\citet{keet2015ontology} annotate the constructs of UML class diagrams, EER, and ORM2 with a shared set of ontological categories to enable comparison and translation across these otherwise incompatible notations.

Across these works, semantic annotation acts as a lightweight, tool-supported layer that complements the metamodel by making the intended role of its elements explicit and machine-usable, without requiring the modeling language itself to be redefined. We build on this same principle, but apply it to a task not addressed by prior work: guiding an LLM through the iterative completion of a partial model, using a small, DSL-independent set of semantic roles which we call a semantic annotation scheme.

\paragraph{\textbf{Domain Model completion.}}
Model completion is the task of extending a partial model with new elements that are properly integrated into it, so that the result is well-formed with respect to its metamodel~\cite{sen2007domain}. The task has since been extended toward automation, generating completions with little or no manual specification effort from the modeler~\cite{sen2010towards}, and toward recommendation, ranking or suggesting candidate completions for the modeler to accept or discard rather than generating a single deterministic result~\cite{kuschke2013recommending}. In domain model completion, the partial model describes an application domain, so well-formedness is not enough: completions must also enrich it with relevant domain-specific vocabulary and concepts. More recently, the recommendation view has been combined with natural-language processing to autocomplete partial domain models directly from textual descriptions~\cite{burgueno2021nlp}, and with few-shot prompting of large language models to automate completion without task-specific training~\cite{chaaben2023towards}.


Beyond filling gaps, completion also involves deciding what to do next during interactive modeling: which element to introduce, where to place it, and how it should connect to existing parts of the model \cite{welter2026nextfocus}. 
\paragraph{\textbf{Generative AI and LLMs.}}
In recent years, the Natural Language Processing (NLP) domain has dramatically evolved, largely driven by the emergence of large language models (LLMs). These models use deep neural networks, typically based on transformer architecture~\cite{vaswani2017attention}, and are pre-trained on vast textual data to capture human-generated knowledge for generative tasks~\cite{ozkaya2023application}. Their human-like outputs stem from learning statistical regularities in language~\cite{hindle2012naturalness}.

LLMs have been used across the software development lifecycle, from requirements engineering~\cite{ronanki2023investigating} to architecture~\cite{ahmad2023towards} and testing~\cite{guilherme2023initial}, as well as for organizing software engineering knowledge~\cite{syriani2024screening}. In some cases, they require \emph{fine-tuning} to adapt to specific tasks~\cite{min2023recent}, although this can be computationally expensive and limited by data availability.

Thanks to large-scale pre-training, \emph{In-Context Learning (ICL)} enables LLMs to adapt at inference time through prompting~\cite{dong2022survey}. Prompting strategies specify task instructions and expected outputs, optionally with examples. Compared to fine-tuning, this lightweight adaptation requires less effort and can be combined with parameter tuning (e.g., temperature, top-$p$) and iterative prompt refinement~\cite{zhang2023instruction, white2023prompt}.

\revadd{A practical distinction concerns the deployment model of LLMs. Commercial models such as GPT-4 \cite{achiam2023gpt} offer strong generative capabilities but require sending data to external cloud services, which raises concerns regarding data privacy, confidentiality, and reproducibility in industrial and research settings. Open-source models such as 
LLaMA~\cite{grattafiori2024llama} and Qwen3~\cite{yang2025qwen3}, which provides a family of models of different sizes, can be deployed locally, giving organizations full control over their data and infrastructure. Although open-source models may have narrower prior knowledge than their commercial counterparts, this gap can be mitigated through explicit guidance mechanisms, as explored in this work.}

\section{Approach}

\label{sec:approach}

Our approach addresses domain model completion by combining three complementary mechanisms. First, it relies on a set of static artifacts that capture both the structural constraints of the target DSL and the semantic roles of its modeling elements. Second, it distinguishes between three types of modeling actions, each requiring a different prompting strategy. Third, it orchestrates these elements through a scoring mechanism that adapts to the evolving model state. The following subsections present an overview of the architecture, the modeling setup and semantic annotation required for each DSL, the orchestration engine that drives candidate selection, and the tool integration into Sirius Web.

\subsection{Overview}
\label{sec:overview}

\definecolor{inputcol}{RGB}{255,230,170}   
\definecolor{outputcol}{RGB}{205,190,245}  
\definecolor{subpcol}{RGB}{230,230,230}    

\definecolor{step1col}{RGB}{250,245,220}   
\definecolor{step2col}{RGB}{255,228,150}  
\definecolor{step3col}{RGB}{255,215,215}   
\definecolor{step4col}{RGB}{220,235,255}

\tikzset{
  myblock/.style = {draw, rounded corners, thick, align=center,
                    minimum width=25mm, minimum height=8mm, fill=white},
  io/.style      = {myblock, fill=inputcol},
  outp/.style    = {myblock, fill=outputcol},
  subp/.style    = {myblock, fill=subpcol},
  line/.style    = {-{Latex[length=2.2mm]}, thick},
  grp/.style     = {rounded corners, draw=black!40, thick}
}

Domain model completion has traditionally been studied for specific modeling formalisms, such as UML class diagrams~\cite{chaaben2023towards} or goal-oriented 
models~\cite{chen2023use}, where completion mechanisms are tightly coupled to a particular language and its constraints. In contrast, our approach is designed to generalize across modeling formalisms, e.g., UML, SysML-V2 with its different versions, etc. 
We realize our approach in \textbf{\tool (\emph{Recommender for AI-assisted Framework for domaIn model Completion})}, an implementation built on top of Sirius Web, an open-source web-based language workbench that enables interactive and incremental modeling. \tooll operationalizes the approach by integrating LLM-generated suggestions into the modeling environment and orchestrating their generation as the model evolves. It iteratively proposes extensions to the model under construction while leveraging the DSL's structural and semantic information to guide and constrain the completion process.


\paragraph{\textbf{Artifacts and preparation.}}
The approach relies on a set of artifacts, summarized in
Table~\ref{tab:rafic-inputs} and illustrated in
Figure~\ref{fig:architecture-overall}. These artifacts are configured once
for each DSL and then reused across modeling sessions. The
\texttt{metamodel} and \texttt{concrete syntax} are provided by the
language designer, as in a standard language workbench setup. The
\texttt{semantic annotation scheme} is a lightweight scheme provided by
\tooll that associates metamodel elements with generic modeling functions
(e.g., process, goal, or resource). The language designer, possibly with
support from domain experts, then defines the \texttt{role-metamodel
mapping} by associating DSL concepts and relations with the corresponding
categories of the annotation scheme.

The metamodel and role-metamodel mapping constitute the static guidance
used during completion. The former captures the structural constraints of
the DSL, while the latter provides semantic guidance by linking its
constructs to their intended modeling roles. Together, they support prompt
instantiation, candidate prioritization, and conformance checking. The
concrete syntax serves a different purpose: it defines how models are
represented and manipulated in the editor and is therefore not included
as prompt content.
\begin{figure*}[t]
\centering
\scriptsize

\begin{minipage}[t]{0.42\textwidth}
\vspace{0pt}
\centering

\renewcommand{\arraystretch}{1.05}
\setlength{\tabcolsep}{3pt}

\begin{tabularx}{\linewidth}{@{}p{0.25\linewidth} p{0.28\linewidth} X@{}}
\hline
\rowcolor{step1col}
\textbf{Artifact} & \textbf{Category} & \textbf{Role} \\
\hline

\rowcolor{inputcol}
Metamodel 
& Static guidance input 
& Defines the language structure and constraints; guides prompts, scoring, and validation. \\
\hline

\rowcolor{inputcol}
\scriptsize Role-Metamodel mapping
& Static guidance input 
& Links metamodel elements  and semantic annotation categories; guides semantic suggestion and prioritization. \\
\hline

\rowcolor{subpcol}
Concrete syntax 
& Static editor configuration 
& Defines how the solution is written and displayed in the editor; not used in prompts. \\
\hline

\rowcolor{subpcol}
Evolving solution 
& Runtime artifact 
& Current model; updated after accepted suggestions or manual edits. \\
\hline

\rowcolor{subpcol}
User feedback 
& Runtime artifact 
& Accept/reject decisions and manual changes. \\
\hline

\rowcolor{outputcol}
Suggestions 
& Runtime output 
& Candidate model extensions proposed at each iteration. \\
\hline
\end{tabularx}

\captionof{table}{Static configuration inputs and runtime artifacts used by \tool.}
\label{tab:rafic-inputs}

\end{minipage}
\hfill
\begin{minipage}[t]{0.55\textwidth}
\vspace{0pt}
\centering

\resizebox{\linewidth}{!}{%
\begin{tikzpicture}[font=\small]

\tikzset{
  state/.style = {
    rectangle,
    rounded corners,
    draw=black!50,
    dashed,
    fill=subpcol,
    minimum width=31mm,
    minimum height=13mm,
    inner ysep=4pt,
    align=center
  }
}

\def\yMid{-1.5}
\def\yUser{-2.65}
\def\yModel{-3.45}

\fill[step1col,draw=black!30,grp] (-1.2,1.1) rectangle (8.5,-4.45);

\draw[rounded corners, draw=black!30, thick]
  (-1.05,0.92) rectangle (2.05,-4.20);
\node[font=\small\bfseries] at (0.5,0.55) {Static setup};

\draw[rounded corners, draw=black!30, thick]
  (2.20,0.92) rectangle (8.33,-4.20);
\node[font=\small\bfseries] at (4.5,0.55) {Dynamic execution};

\node[io, minimum width=22mm] (mm) at (0.2,-0.5) {Metamodel};
\node[io, minimum width=22mm] (map) at (0.2,\yMid) {\scriptsize Role-Metamodel\\mapping};
\node[subp, minimum width=24mm] (CS) at (0.2,-2.5) 
  {Concrete syntax\\\footnotesize editor notation};

\node[subp, minimum width=30mm, minimum height=12mm] (ws) at (3.95,\yMid)
  {Modeling workspace\\\footnotesize (editor + orchestrator)};

\node[state] (model) at (3.95,\yModel)
  {Evolving solution\\\footnotesize (initially empty/partial)};

\node[outp, minimum width=20mm] (sugg) at (7.2,\yMid)
  {Suggestions};

\node[myblock, minimum width=20mm, minimum height=10mm] (user) at (7.2,\yUser)
  {User:\\ accept / reject};

\draw[line] (mm.east) -- (ws.west);
\draw[line] (map.east) -- (ws.west);
\draw[dashed, line] (CS.east) -- node[below, sloped, font=\scriptsize] 
  {representation} (ws.west);
\draw[line] (ws.east) -- (sugg.west);
\draw[line] (sugg.south) -- (user.north);

\draw[line] (user.west) -- ++(-0.45,0) |- ([yshift=-2mm]ws.east);

\draw[line] ([xshift=-5mm]ws.south) -- ([xshift=-5mm]model.north);
\draw[line] ([xshift=5mm]model.north) -- ([xshift=5mm]ws.south);

\end{tikzpicture}
}

\captionof{figure}{Overview of RAFiC showing the interaction between static knowledge, dynamic orchestration, and user feedback in the iterative completion loop.}
\label{fig:architecture-overall}

\end{minipage}

\end{figure*}

\paragraph{\textbf{Iterative completion loop.}}
During execution, \tooll operates over an evolving solution, which may initially be empty or partial and is updated whenever the user accepts a suggestion or manually adds an element. At each iteration, the orchestration engine analyzes the current model, identifies structural and semantic gaps, and selects the most relevant candidate completions. It then instantiates a prompt tailored to the selected gap and sends it to the LLM. The generated suggestions are presented to the user, who may accept or reject them. Accepted suggestions update the model; rejected suggestions are discarded but may influence subsequent scoring through updated context. To keep suggestions focused, the engine distinguishes between three types of modeling actions: architectural (ARCH), for introducing new concepts; relational (REL), for connecting existing ones; and characteristic (CHAR), for enriching them with attributes, labels, or constraints. The same three types also drive prompt instantiation, each with its own family of templates (Section~\ref{prompt_instantiation}).


\subsection{Modeling Setup }
\label{sec:setup}
This section describes the preparation required to apply \tooll to a new DSL. We first introduce a running example, then detail the required artifacts: metamodel, concrete syntax, semantic annotation scheme, and role-metamodel mapping.

\begin{figure*}[h]
  \centering

  \begin{subfigure}[t]{0.315\textwidth}
    \centering
    \fcolorbox{black!10}{white}{%
      \includegraphics[width=\linewidth]{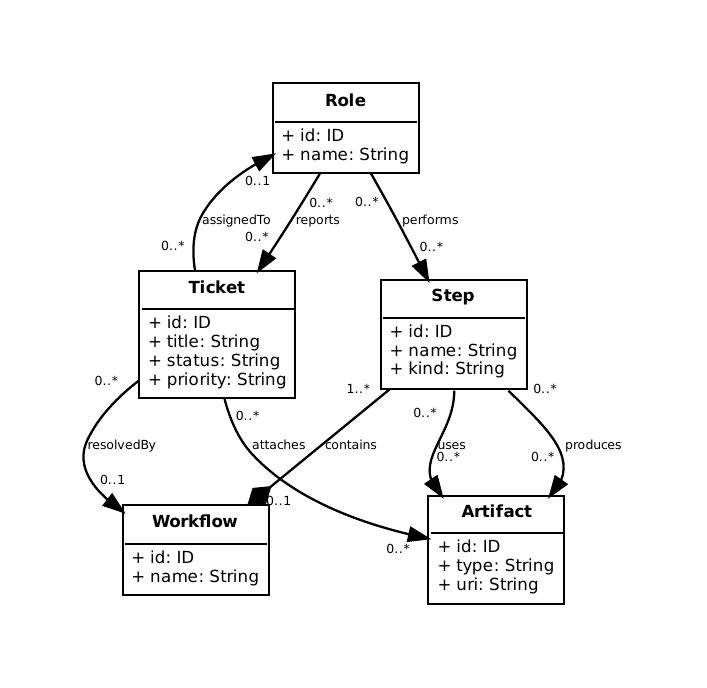}%
    }
    \caption{\footnotesize Metamodel}
    \label{fig:ticketing-metamodel}
  \end{subfigure}\hfill
  \begin{subfigure}[t]{0.20\textwidth}
    \centering
    \fcolorbox{black!10}{white}{%
      \includegraphics[width=0.75\linewidth]{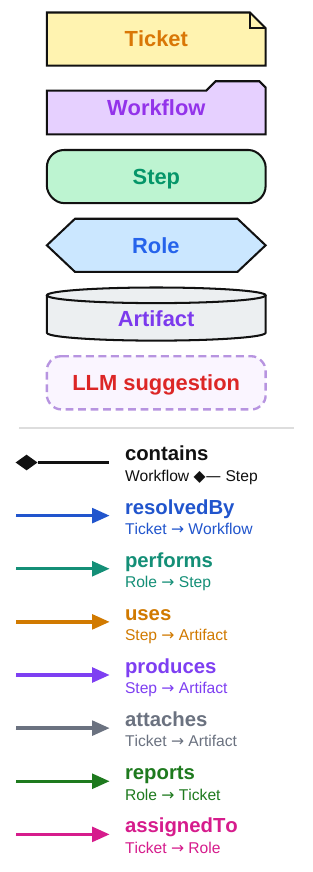}%
    }
    \caption{\footnotesize Concrete Syntax (CS)}
    \label{fig:ticketing-cs}
  \end{subfigure}\hfill
  \begin{subfigure}[t]{0.45\textwidth}
    \centering
    \fcolorbox{black!10}{white}{%
      \includegraphics[width=\linewidth]{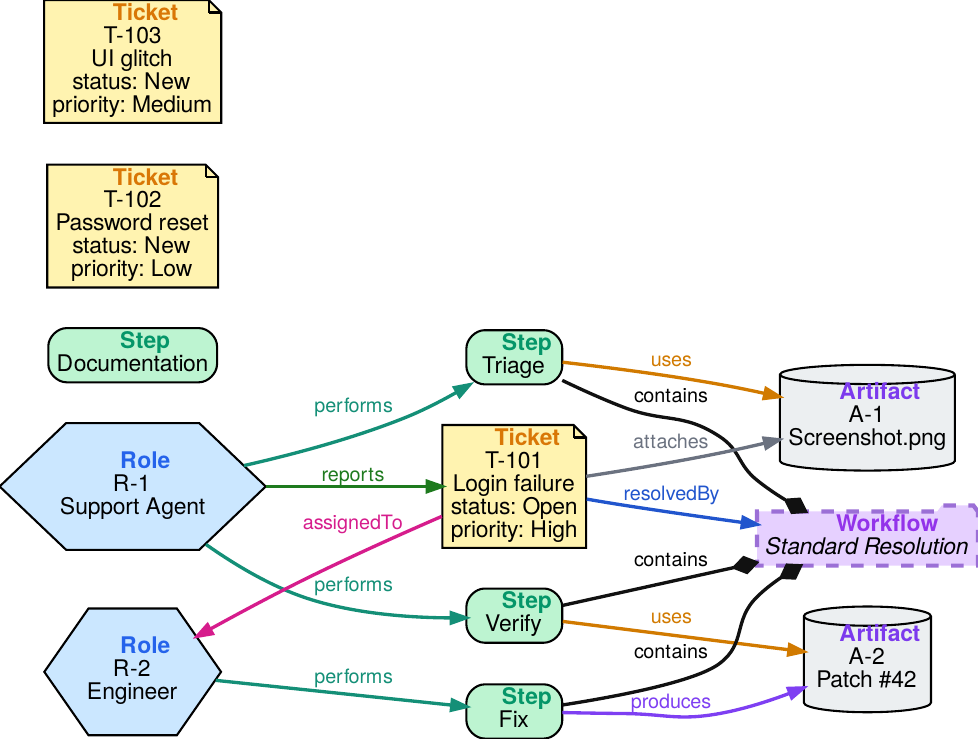}%
    }
    \caption{\footnotesize Instance or model under construction}
    \label{fig:ticketing-instance}
  \end{subfigure}

  \caption{Ticketing DSL: (a) metamodel, (b) concrete syntax, and (c) example instance.}
  \label{fig:ticketing-triptych}
\end{figure*}

\paragraph{Running example:}
We use a compact \emph{``Ticketing DSL''} to model how support requests are taken from intake to resolution (Figs.~\ref{fig:ticketing-metamodel}, \ref{fig:ticketing-cs}, and \ref{fig:ticketing-instance}), and refer back to it throughout the following subsections. Tickets are progressed through a structured workflow of steps, carried out by responsible roles and accompanied by evidence or outputs. Fig.~\ref{fig:ticketing-metamodel} shows a minimal Ticketing metamodel in which a Workflow contains Steps, Roles perform Steps, a Ticket can be resolved by a Workflow, and Steps may use or produce Artifacts. A Ticket is also reported by a Role, may be assigned to a Role, and may attach Artifacts as evidence. Fig.~\ref{fig:ticketing-cs} gives one notation per metamodel class and reference. Fig.~\ref{fig:ticketing-instance} shows a model under construction: T-102, T-103 and the step \textsf{Documentation} are not yet connected, and the highlighted Workflow name, \textsf{Standard Resolution}, is a suggestion accepted by the user.

\subsubsection{\textbf{Metamodel and concrete syntax.}} 
Before using \tool,  the language designer defines the DSL-specific artifacts introduced in Section~\ref{sec:overview}, starting with the DSL metamodel and its concrete syntax. The metamodel specifies the abstract syntax of the language, including modeling concepts, relations, containment rules, and constraints. Following standard practice in metamodeling frameworks such as Ecore, the metamodel itself is expressed as a class diagram containing classes, attributes, and references, independently of whichever concrete syntax the DSL uses to display its instance models.
 \tooll uses the metamodel artifacts as a structural source for prompt instantiation, candidate prioritization, and conformance checking. The concrete syntax defines how models are written, displayed, and manipulated in the editor. It supports user interaction in the modeling workbench, but is not used as prompt content for generating suggestions.

\subsubsection{\textbf{Semantic-annotation scheme definition.}}
\revadd{
The semantic annotation scheme is defined once as part of the \tooll framework and can be reused across different DSLs. Only the role-metamodel mapping, described below, is DSL-specific.
The design of this scheme was motivated by the observation that many modeling languages, despite their syntactic differences, share recurring conceptual roles: elements that represent processes, resources, actors, goals, constraints, or structural containers. Several of these roles have direct precedent in the literature; for instance, \emph{Process} corresponds to a formal category in the foundational ontology DOLCE~\cite{gangemi2002sweet}, \emph{Role} is a signature category of the Unified Foundational Ontology (UFO)~\cite{guizzardi2005ontology,guizzardi2022ufo}, and \emph{Thing} is a primitive construct of the Bunge--Wand--Weber (BWW) ontology for information systems~\cite{wand1990ontological}. 
Importantly, the purpose of this scheme is not to define a universal formal semantics for modeling languages. Rather, it provides lightweight semantic hints that help the LLM interpret the intended role of a language construct and connect it to the general meaning of domain concepts acquired during pretraining. By making this connection explicit, the annotations help bridge the gap between the domain knowledge that an LLM may possess and the language-specific constructs used to express that knowledge in a model.
}

\revadd{
We defined the semantic-annotation scheme using an iterative refinement process. We began by collecting metamodels from several modeling languages, ranging from general-purpose notations such as UML class diagrams, SysML, goal models, ER diagrams, and state machines, to more narrowly scoped domain-specific languages studied in prior work, such as WebDSL for web application development~\cite{visser2008webdsl} and a visual DSL for elevator controllers~\cite{wienands2009anatomy}. Then, we identified recurring modeling roles independently of their DSL-specific names.
}
\revadd{
For instance, concepts that described elements acting or unfolding over time, such as activities and behaviors were grouped under a single \emph{Process} role; resource- and structure-oriented concepts were grouped under \emph{Thing}; agent-oriented concepts under \emph{Role}; and objective- or requirement-oriented concepts under \emph{Goal}. Concepts that did not recur across the DSLs we examined, or that were too fine-grained to be useful for prompt construction, were dropped. We repeated this process iteratively, refining the candidate set at each pass, until the retained categories, such as \emph{Process}, \emph{Thing}, \emph{Role}, and \emph{Goal}, consistently captured the concepts we encountered. Relations were derived the same way, keeping only those that expressed a recurring semantic link useful for completion, such as composition, participation, use/production, achievement, or constraint.}\revadd{ The resulting scheme was collaboratively reviewed by the four authors, who checked whether its categories and relations could capture the intended roles of the DSL concepts considered in our case studies and additional DSL examples explored during design.
}

\begin{figure}[h]
    \centering
    \includegraphics[width=1\linewidth]{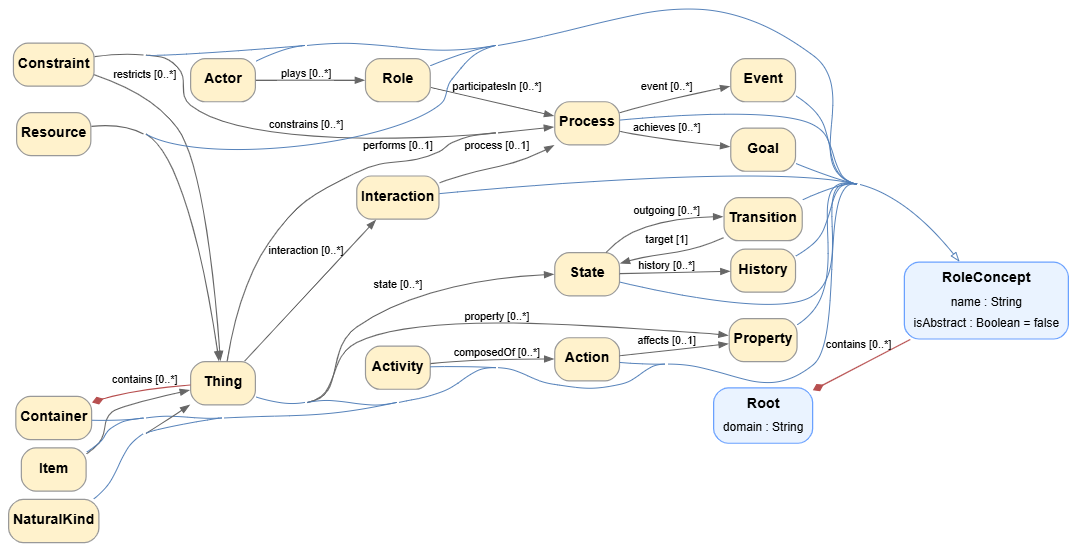}
    \caption{Semantic annotation scheme used to annotate metamodel elements with their intended modeling role.}
    \label{fig:ontology-metamodel}
\end{figure}

Figure~\ref{fig:ontology-metamodel} shows the resulting structure. A \emph{Root} element, with a \texttt{domain} attribute, contains all role concepts, and every category specializes the abstract type \emph{RoleConcept}. Its \texttt{isAbstract} boolean attribute lets a category be kept abstract or refined into a concrete one depending on the target DSL; for example, in SysML~v2, some concepts are kept abstract to reflect the distinction between definitions and usages. The scheme is thus a flexible semantic framework rather than a fixed hierarchy.

The scheme acts as a lightweight semantic annotation layer: it enriches the prompts given to the LLM, supports candidate prioritization, and guides action selection, allowing \tooll to communicate the intended modeling role of a DSL element rather than relying only on its name. For example, if a DSL element \textsf{Ticket} is mapped to \emph{Goal}, the prompt can frame it as an objective or issue to be resolved; if \textsf{Step} is mapped to \emph{Process}, the prompt can frame it as an action contributing to that resolution.

\subsubsection{\textbf{Role-metamodel mapping.}}

After the semantic annotation scheme is available, the language designer defines a \emph{role-metamodel mapping}, which links metamodel elements, and when relevant relation pairs, to concepts and relations from the annotation scheme. As noted in the overview section~\ref{sec:overview}, this mapping is created once per DSL during language setup and reused across modeling sessions. It can be performed by the language designer alone, or with support from domain experts when the semantics of the DSL concepts require domain-specific interpretation.

The mapping follows a simple procedure. First, each metamodel class is assigned to the closest annotation category according to its role in the model. For instance,
classes representing actions, workflows, behaviors, or states of progress are mapped to \emph{Process}; classes representing resources, documents, data, or physical objects are mapped to \emph{Thing}; classes representing actors, responsibilities, or organizational units are mapped to \emph{Role}; and classes representing objectives, requirements, issues, or desired outcomes are mapped to \emph{Goal}. Second, metamodel references are mapped to annotation  relations according to their intended meaning, such as participation, use/production, composition, achievement, or constraint.

\paragraph{Example applied to the Ticketing metamodel.} 
We denote by $\mathcal{M}$ the role-metamodel mapping, a function from metamodel symbols to annotation-scheme categories  and relations: $\mathcal{M}:\Sigma_{\mathsf{meta}} \to\Sigma_{\mathit{role}}$.

We overload $\mathcal{M}$ for both concepts and relations, preserving the argument structure: binary relations map to binary annotation relations, and unary concepts map to unary annotation  categories. For the Ticketing DSL example, all mapped annotation  elements are concrete, i.e., their attribute \texttt{isAbstract} is set to \texttt{false}. The mapping used in the running example is summarized below:

\setlength{\tabcolsep}{2.5pt}
\renewcommand{\arraystretch}{0.95}

\noindent\footnotesize
\begin{tabular}{@{}r@{\,$\mapsto$\,}l @{\qquad} r@{\,$\mapsto$\,}l@{}}
\(\M(\mathsf{Workflow})\) & \(\mathit{Process}\)
& \(\M(\mathsf{contains})\) & \(\mathit{structural\,composition}\) \\
\(\M(\mathsf{Step})\)     & \(\mathit{Process}\)
& \(\M(\mathsf{performs})\) & \(\mathit{participation}\) \\
\(\M(\mathsf{Role})\)     & \(\mathit{Role}\)
& \(\M(\mathsf{uses})\)     & \(\mathit{use/production}\) \\
\(\M(\mathsf{Artifact})\) & \(\mathit{Thing}\)
& \(\M(\mathsf{produces})\) & \(\mathit{use/production}\) \\
\(\M(\mathsf{Ticket})\)   & \(\mathit{Goal}\)
& \(\M(\mathsf{resolvedBy})\) & \(\mathit{achievement}\) \\
\end{tabular}

\normalsize

The role-metamodel mapping  narrows the completion space and specializes prompts for likely next steps. The orchestration engine, described next, uses these artifacts at each iteration to score candidates and select the most relevant completions. References without a mapped role (e.g., \textsf{reports}) are handled by the structural scores only.

\subsection{Orchestration Engine}
\label{sec:orchestration}
Let $M_t$ denote the model at iteration $t$, where $M_0$ is the initial partial model. At each iteration, the system constructs a set of candidate completions $\mathcal{C}(M_t)$, where each candidate  $c \in \mathcal{C}(M_t)$ corresponds to a possible modeling action. Each candidate is assigned a score $S(c)$ reflecting its relevance in the current context. The engine selects the top-$K$ candidates and instantiates them into prompts. Accepted candidates update the model: $M_{t+1} = M_t \cup \{c\}$.  The remainder of this section details how candidates are scored (Section~\ref{scoring}), how the top-$K$ are dispatched (Section~\ref{dispatch}), and how prompts are constructed from three distinct action families (Section~\ref{prompt_instantiation}).

\subsubsection{\textbf{Candidate scoring.}}
\label{scoring}
Each candidate receives a composite score:
\[
S(c) = S_{\mathrm{metamodel}}(c) + 
S_{\mathrm{canvas}}(c) + S_{\mathrm{semantic}}(c)
\]
The scoring components use fixed heuristic weights in our prototype. They are not claimed to be optimal; rather, they encode simple modeling priorities, such as introducing structurally central elements, repairing missing relations, and favoring semantically  meaningful completions. We describe each component below.

\paragraph{(1) $S_{\mathrm{metamodel}}$: structural prior 
and entrypoints.} 
The prior $S_{\mathrm{metamodel}}(c)$ ranks candidates using only the metamodel, independently of the current canvas or role annotations. It highlights actions that the language makes structurally important, such as introducing central elements, establishing required relations, or completing mandatory characteristics. We compute this score using simple fixed heuristics (Table~\ref{tab:heuristics}), which provide a first ranking before any context-specific adjustment.
It also identifies a small set of \emph{entrypoints}, $\mathcal{E}_{\text{entry}} = \operatorname{TopK}_{c} S_{\mathrm{metamodel}}(c)$, which correspond to modeling actions that should be introduced early. If the canvas still lacks enough instances of these entrypoints, selection is temporarily biased toward them until this initial backbone is established.

In the Ticketing running example, the resulting scores rank \textsf{Ticket} highest (Table~\ref{tab:topk}). This makes it natural to start from \textsf{Ticket}s, before refining \textsf{Step}s, \textsf{Artifact}s, \textsf{Role}s, and \textsf{Workflow}s.
\begin{table}[H]
\centering
\begin{minipage}[t]{0.42\columnwidth}
\centering
\footnotesize
\begin{tabular}{@{}ll@{}}
\toprule
\multicolumn{2}{c}{\textbf{Entrypoint scoring heuristics}} \\
\midrule
Containment & $+5$ per containment \\
Outgoing ref. & $+2$ each \\
Incoming ref. & $+1$ each \\
``Many'' rel. & $+3$ each \\
\bottomrule
\end{tabular}
\caption{Scoring rules.}
\label{tab:heuristics}
\end{minipage}\hfill
\begin{minipage}[t]{0.54\columnwidth}
\centering
\footnotesize
\begin{tabular}{@{}lcccc|c@{}}
\toprule
Entity & Cont. & In & Out & Many & \textbf{Total} \\
\midrule
Ticket    & 0  & +1 & +6 & +12 & \textbf{19} \\
Step      & 0  & +1 & +4 & +12 & \textbf{17} \\
Artifact  & 0  & +3 & 0  & +9  & \textbf{12} \\
Role      & 0  & +1 & +4 & +6  & \textbf{11} \\
Workflow  & +5 & +1 & 0  & 0   & \textbf{6}  \\
\bottomrule
\end{tabular}
\caption{Top-\(K\) entrypoint elements for the Ticketing example, ranked by \(S_{\mathrm{metamodel}}\).}
\label{tab:topk}
\end{minipage}
\end{table}
\paragraph{(2) $S_{\mathrm{canvas}}$: canvas-aware adjustment.}
As the model evolves, $S_{\mathrm{canvas}}(c)$ adjusts this initial ranking according to what is already present on the canvas. It raises candidates whose surrounding structural obligations are still unmet, such as missing relations, incomplete local neighborhoods, or pending characteristics. Prioritization therefore reflects both the metamodel and the current gaps of the partial model.

\paragraph{(3) $S_{\mathrm{semantic}}$: Semantic-aware prioritization.}
$S_{\mathrm{semantic}}(c)$ uses the role-metamodel mapping $\mathcal{M}$  to favor semantically meaningful completions. It promotes candidates that match important semantic scheme patterns, such as \emph{Process}-\emph{Role}, 
\emph{Process}-\emph{Thing}, and \emph{Goal}-\emph{Process}. Using these patterns, \tooll detects missing relations and proposes semantically appropriate completions. For example, if a Step lacks a resource, the system may propose:
$\textsf{Step} \xrightarrow{\mathit{uses}} \textsf{Artifact}.$

If a Ticket is unresolved, the system may propose: $\textsf{Ticket} \xrightarrow{\mathit{resolvedBy}} \textsf{Workflow}$  and then suggest an initial Step toward resolution. By contrast, purely structural ARCH suggestions that do not support participation, resource use, or goal achievement are down-weighted.
\\ Table~\ref{tab:unified-next-focus-high} illustrates the combined effect of $S_{\mathrm{canvas}}$ and $S_{\mathrm{semantic}}$ as a ranked list of focus patterns. 
\newline
\newcolumntype{Y}{>{\raggedright\arraybackslash}X}

\begin{table}[h]
\scriptsize
\setlength{\tabcolsep}{2.2pt}
\renewcommand{\arraystretch}{0.86}

\begin{tabularx}{\linewidth}{
@{} c c
>{\centering\arraybackslash}p{0.055\linewidth}
>{\hsize=0.95\hsize}Y
>{\hsize=0.80\hsize}Y
>{\hsize=1.25\hsize}Y
>{\raggedleft\arraybackslash}p{0.055\linewidth}
@{}}
\toprule
\textbf{It.} & \textbf{Rk.} & \textbf{Type} &
\textbf{Mini-task} &
\textbf{Elements} &
\textbf{Reason} &
\textbf{$S$} \\
\midrule

\multicolumn{7}{@{}p{\linewidth}@{}}{
\textbf{Iteration 1.}
Canvas: \textit{Ticket} TK=T-101 ``Login failure''; unnamed \textit{Workflow} WF containing
\{TR=Triage, VF=Verify, FX=Fix\}; no Roles; no Artifacts.
} \\
\midrule

1 & 1 & \textbf{REL}  & Wire performer(s) to process steps & \textit{new Role} $\to$ TR (\textsf{performs}) & [Wire] steps lack participants; [Sem] Process--Role; needs Rk.~2 & 90 \\
\rowcolor{accepted}
1 & 2 & \textbf{ARCH} & Introduce missing actor/role & SA:\textsf{Role} & [AfterE] participant gap; boosts \textbf{REL} {\scriptsize(accepted)} & 88 \\
1 & 3 & \textbf{REL}  & Wire step to its output & FX $\to$ \textit{new Artifact} (\textsf{produces}) & [Wire] output missing; [Sem] use/production & 86 \\
1 & 4 & \textbf{ARCH} & Provision output artifact(s) & PT:\textsf{Artifact} & [Spawn] to satisfy Rk.~3; prefer reuse & 84 \\
1 & 5 & \textbf{REL}  & Wire step to evidence & TR $\to$ \textit{new Artifact} (\textsf{uses}) & [Wire] evidence missing; [Sem] use/production & 79 \\
1 & 6 & \textbf{ARCH} & Provision evidence artifact(s) & SC:\textsf{Artifact} & [Spawn] for TR; complements Rk.~5 & 78 \\
1 & 7 & \textbf{REL}  & Bind goal to execution structure & TK $\to$ WF (\textsf{resolvedBy}) & [Sem] Goal--Process; achievement path & 71 \\
1 & 8 & \textbf{CHAR} & Add essential attributes to central items & TK.\texttt{priority} & [AfterE] few attrs & 66 \\

\midrule
\multicolumn{7}{@{}p{\linewidth}@{}}{
\textbf{Iteration 2.}
Canvas change: SA=\textsf{Support Agent} \emph{accepted}. Other items still pending.
} \\
\midrule

\rowcolor{chgup}
2 & 1 & \textbf{REL}  & Wire accepted role to key step & SA $\to$ TR (\textsf{performs}) & [Wire] now possible; [Sem] Process--Role & $97{\scriptsize(+7)}$ \\
\rowcolor{chgup}
2 & 2 & \textbf{ARCH} & Provision output artifact(s) & PT:\textsf{Artifact} & [Spawn] needed by Rk.~4 & $89{\scriptsize(+5)}$ \\
\rowcolor{chgnew}
2 & 3 & \textbf{REL}  & Extend performer wiring across related steps & SA $\to$ VF (\textsf{performs}) & [Wire] sibling step; [Inc] popular target {\scriptsize[new]} & 88 \\
\rowcolor{chgsame}
2 & 4 & \textbf{REL}  & Wire step to its output & FX $\to$ PT (\textsf{produces}) & [Wire] output gap; [Many] list-like & $86{\scriptsize(+0)}$ \\
\rowcolor{chgup}
2 & 5 & \textbf{REL}  & Wire step to evidence & TR $\to$ SC (\textsf{uses}) & [Wire] evidence gap & $81{\scriptsize(+2)}$ \\
\rowcolor{chgup}
2 & 6 & \textbf{ARCH} & Provision evidence artifact(s) & SC:\textsf{Artifact} & [Spawn] complements Rk.~5 & $81{\scriptsize(+3)}$ \\
\rowcolor{chgup}
2 & 7 & \textbf{REL}  & Bind goal to execution structure & TK $\to$ WF (\textsf{resolvedBy}) & [Sem] Goal--Process & $74{\scriptsize(+3)}$ \\
\rowcolor{chgnew}
2 & 8 & \textbf{CHAR} & Add operational attribute to step & FX.\texttt{kind} & [AfterE] few attrs on step {\scriptsize[new]} & 73 \\

\bottomrule
\end{tabularx}

\vspace{1pt}
\begin{minipage}{\linewidth}
\tiny
\textbf{Legend:}
\fcolorbox{black!5}{chgup}{\strut\hspace{6pt}} increase;
\fcolorbox{black!5}{chgnew}{\strut\hspace{6pt}} new;
\fcolorbox{black!5}{chgsame}{\strut\hspace{6pt}} unchanged;
\fcolorbox{black!5}{accepted}{\strut\hspace{6pt}} accepted.
Items falling below rank 8 (e.g., TK.\texttt{priority}, 66) are omitted.
\textbf{Abbrev.:}
TK=\textsf{T-101}, WF=\textsf{Workflow}, TR=\textsf{Triage}, VF=\textsf{Verify}, FX=\textsf{Fix},
SA=\textsf{Support Agent}, PT=\textsf{Patch \#42}, SC=\textsf{Screenshot.png}.
\textbf{Types:}
\textbf{ARCH}=architectural, \textbf{REL}=relational, \textbf{CHAR}=characteristic.
\textbf{Tags:}
[Many]=multiplicity, [Inc]=incoming-popularity,
[AfterE]=attributes-after-entity, [Wire]=wire-after-shape,
[Spawn]=spawn-and-wire, [Sem]=semantic affinity.
\end{minipage}
\caption{Unified ranking of the next modeling operation on the running example: two-iteration snapshot leading to the model of Fig.~\ref{fig:ticketing-instance}.}
\label{tab:unified-next-focus-high}

\end{table}
Iteration~2 shows how accepting one item in Iteration~1 reshapes the pool and raises the priority of the most consequential follow-up actions.

\subsubsection{\textbf{Parallel dispatch.}}
\label{dispatch}
The engine selects the top-$K$ focus elements and dispatches them, potentially in parallel. In practice, $K$ is kept small to limit unnecessary LLM calls, reduce computation time, and avoid overwhelming the user with too many unfocused suggestions. Each candidate then goes through prompt instantiation, generation, and post-processing. The resulting completions are presented to the user, who remains in control by accepting or rejecting the suggestions.

\subsubsection{\textbf{Prompt instantiation}}
\label{prompt_instantiation}
A crucial design decision is to distinguish between three focus categories, each corresponding to a family of prompts. The \textbf{ARCHitectural} focus (ARCH) proposes new domain concepts such as entities, processes, or components. The \textbf{RELational} focus (REL) proposes links between concepts, such as containment, association, flow, or dependency. The \textbf{CHARacteristic} focus (CHAR) enriches existing concepts with attributes, roles, labels, or constraints. For graphical modeling languages, these naturally map to adding or refining nodes (ARCH), edges (REL), or annotations (CHAR). In practice, the engine selects one of three template-based prompts, instantiates it with structural and semantic guidance, and constrains the output to short labels. All templates share the same structure: (i) a fixed system instruction enforcing short label generation, (ii) a user prompt instantiated with contextual placeholders, and (iii) a rigid output format for lightweight parsing. 
The instantiated placeholders include the domain, focus type, partial instance context, optional semantic hint, and optional co-occurrence terms. All templates enforce the same response constraints: labels must be short, unique, explanation-free, and returned in a fixed parsable format.

\begin{table}[H]
\centering
\scriptsize
\setlength{\tabcolsep}{3pt}
\renewcommand{\arraystretch}{1.05}
\caption{Prompt template families used in the approach.}
\label{tab:prompt-templates}
\begin{tabularx}{\columnwidth}{@{}lXl@{}}
\toprule
\textbf{Template} & \textbf{Purpose and instantiated context} & \textbf{Output} \\
\midrule
\textbf{ARCH} & Proposes candidate names for a focal element type using domain, local instance context, optional semantic hint, known facts, and co-occurrence cues. & 1--3  labels \\
\textbf{REL} & Proposes concise relation labels between a focal element and a trigger element using role phrase, trigger, local context, and optional semantic hint. & 1--3 labels \\
\textbf{CHAR} & Proposes characteristic labels (e.g., attributes, roles, states) for a focal element using relation context, trigger, and optional semantic hints. & 1--3 labels \\
\bottomrule
\end{tabularx}
\end{table}

In addition, the prompts are influenced by the role-metamodel mapping. For each
semantic scheme category, a small library of helper templates is defined in
advance. For example, if a Ticket, mapped to the scheme category \emph{Goal},
lacks a resolution path, the system selects the corresponding Goal helper and
adds a semantic hint to the prompt: ``This Ticket represents a goal to be
achieved; propose a Workflow that achieves it.''

\begin{center}
\vspace{4pt}
\begin{tikzpicture}
\node[
draw,
rounded corners,
fill=gray!10,
inner sep=8pt,
text width=0.9\linewidth
]{
\noindent\textbf{Example instantiation (REL, Ticketing).}
\textit{system: You are a precise naming assistant. Follow the
output format exactly. Do not add explanations.}\\
\textit{User: In the ticketing context, consider a Ticket ``Login failure''.
This Ticket represents a goal to be achieved. It is resolved by a Workflow to
be named. This Workflow is composed of the Steps \{Triage, Verify, Fix\}. \\
Suggest up to three names for this Workflow.
Rules: \\
- Single tokens only (A-Z, a-z, 0-9, hyphen, underscore, slash). No spaces.\\
- Each label >= 3 chars. NOT Camel Case. No duplicates. No extra text.}};
\end{tikzpicture}
\end{center}

After post-processing, the generated list is:
\texttt{Standard Resolution, T-101 Workflow, Login Issue Remediation}.
In this case, \texttt{Standard Resolution} is the most suitable choice, as it remains general enough to cover fixes beyond the specific login issue.


\subsection{Tool integration}
\label{sec:Tool}

We implemented our approach \tooll as an open-source tool\footnote{https://anonymous.4open.science/r/Modeling-Aid/README.md}.  We provide in the  repository's README  simplified demonstration videos illustrating both the language-design setup and the recommendation workflow. The tool is composed of two main parts: (i) a backend that orchestrates interactions with the LLM, and (ii) a client integrated into the modeling environment. 

For the  LLM server part, we implemented a dedicated backend responsible for the orchestration between the modeling environment and the language model.
The server exposes a set of REST endpoints to receive model, metamodel, and semantic  data, and to return structured completion recommendations.
It performs  context preparation, and dynamic prompt generation by combining information from the metamodel and semantic  scheme to ensure that LLM prompts and responses remain both syntactically valid and semantically consistent.
The server also maintains a trace of all interactions (inputs, prompts, and decisions) to support the reproducibility and evaluation of the experiment.
Finally, it handles feedback from the client (accepted or rejected suggestions), to inform subsequent recommendations. The client side of \tooll is integrated into  Sirius Web~\footnote{\url{https://github.com/eclipse-sirius/sirius-web/}}, an open-source web-based language workbench that enables the development of graphical domain-specific modeling languages and their associated modeling workbenches directly from a web browser~\cite{giraudet2024sirius}. 
It is developed using Java, Spring Boot, and EMF on the server side, and TypeScript and React on the client side, with GraphQL for communication. 
The platform provides two main meta-languages: the Domain DSL for defining abstract syntax and the View DSL for defining concrete syntax with six default representation types (diagram, form, tree, table, Gantt, and kanban). 
It is domain-agnostic and primarily targets domain experts who need to build graphical DSLs for their specific concerns.
In addition to developing DSLs, Sirius Web is also the platform on which produced modeling workbenches operate.
To separate these different concerns, Sirius Web proposes two types of project: \emph{studio projects} for DSL development; and \emph{basic projects} to use developed DSLs.
Finally, Sirius Web proposes a complete API on which our implementation entirely relies.

First, we added to Sirius Web the metamodel shown in Figure \ref{fig:ontology-metamodel} which we implemented using Ecore, and configured Sirius Web to allow its instantiation only within studio projects.
To map a semantic role concept to a DSL abstract syntax concept, the language designer simply sets the \emph{name} attribute to the corresponding abstract syntax concept name.

\begin{figure}[t]
    \centering
    {\setlength{\fboxsep}{0pt}\setlength{\fboxrule}{0.5pt}%
    \fcolorbox{gray!50}{white}{\includegraphics[width=\dimexpr0.8\linewidth-1pt\relax]{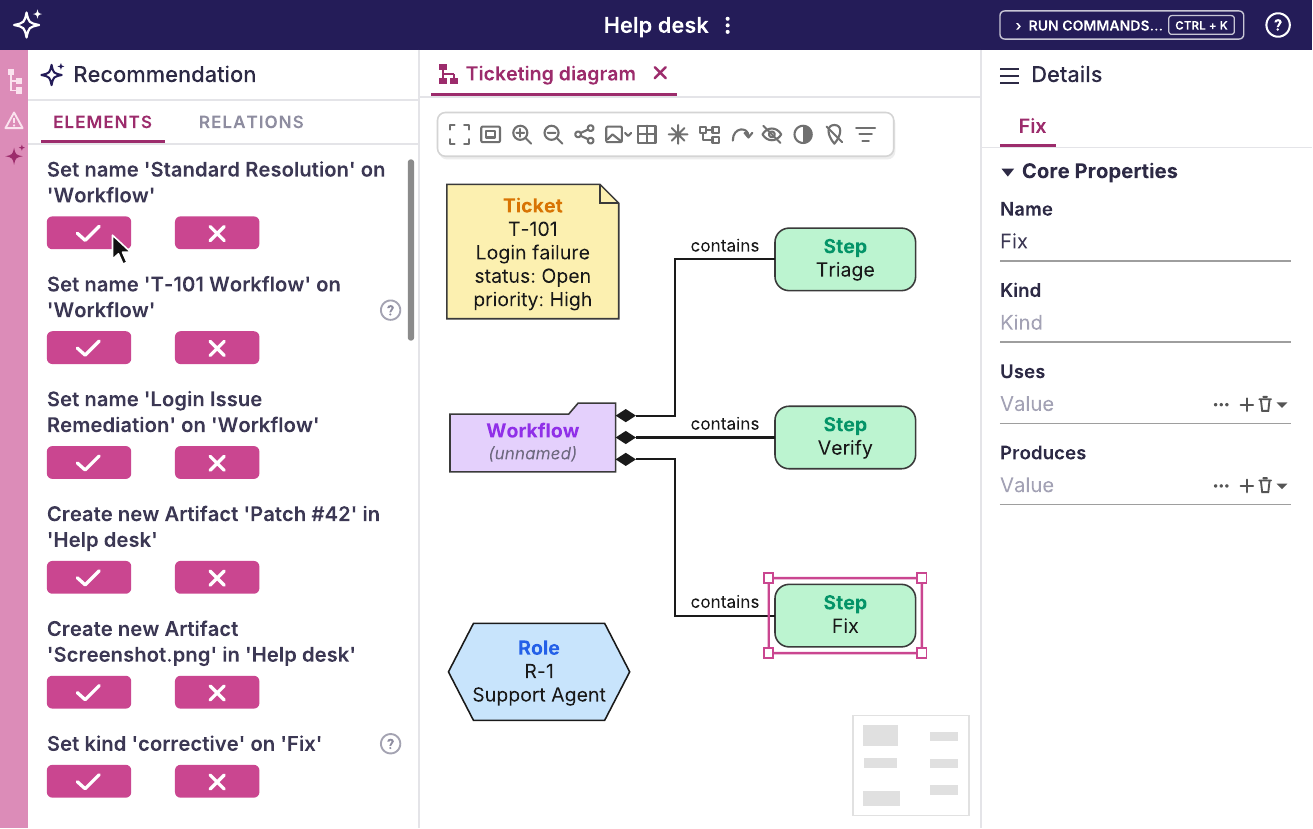}}}
    \caption{\tooll in Sirius Web on the Ticketing example: recommendations (left), model under construction (center), and properties of the selected step (right).}
    \label{fig:rafic-tool}
\end{figure}

On the language user side, \tooll monitors model changes and sends the current model together with its language artifacts to the LLM server when needed. 
The LLM server then pushes recommendations to Sirius Web, which displays them in a dedicated view on the left side of the application, as shown in Figure~\ref{fig:rafic-tool} for the running example.

We consider two kinds of recommendation, each having its own tab in the view: element recommendations (ARCH and CHAR, \emph{i.e.,} new model objects or characteristics); and relation recommendations (REL, connecting existing model objects).
Recommendations are ordered in the view by their frequency, i.e., the number of iterations in which the orchestration engine selected them among the top-$K$ candidates as the canvas evolved. Each recommendation also indicates the model elements to which it applies if accepted (e.g., the Step on which a suggested \texttt{kind} value is set). 
From this view, the language user can accept or reject each recommendation.
Accepting it adds the recommended element or relation to the model; in both cases, the decision is sent back to the LLM server to inform subsequent recommendations. Finally, the recommendations are also displayed in the contextual palette of the elements to which they apply.

\section{Evaluation Methodology}
\label{sec:evaluation}
To assess the effectiveness of our approach, we designed an offline evaluation protocol that isolates the contribution of each guidance mechanism through controlled comparisons and ablations. 
\subsection{Research Questions}

More concretely, we investigate the following three research questions (RQs):

\begin{itemize}
    \item \textbf{RQ1}: How much does \tooll improve over an interactive LLM-based completion baseline that does not incorporate DSL syntax or semantics?
    
    \item \textbf{RQ2}: How much does iterative, context-aware guidance in \tooll improve over an interactive LLM-based approach where DSL syntax and semantics are provided only once upfront?
    
    \item \textbf{RQ3}: What is the contribution of the semantic role guidance within \tool, as assessed through an ablation study?
\end{itemize}

\subsection{Experimental Setup}

\usetikzlibrary{positioning,fit,arrows.meta,backgrounds}
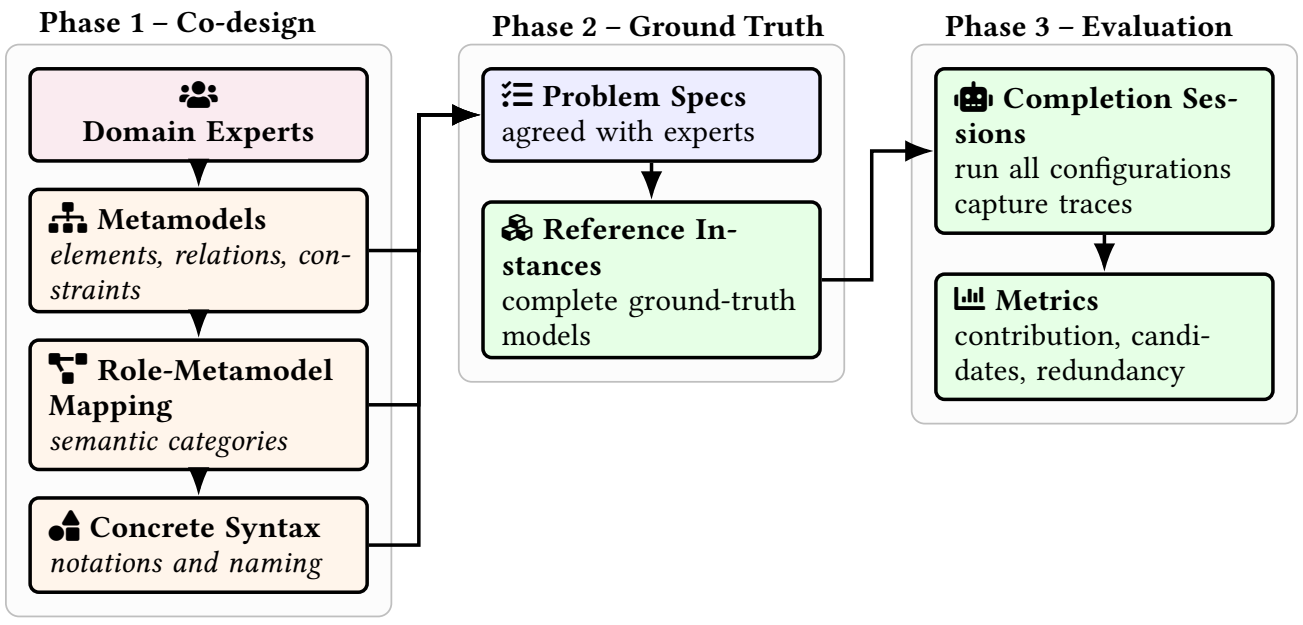
\begin{figure}[H]
\centering
\resizebox{0.62\textwidth}{!}{%
\begin{tikzpicture}[
  font=\scriptsize,
  >=Latex,
  node distance=2mm and 5mm,
  lane/.style={
    draw=black!25,
    rounded corners=3pt,
    inner sep=1.8mm,
    fill=black!1
  },
  title/.style={anchor=west,font=\bfseries\scriptsize},
  iconbox/.style={
    draw,
    rounded corners=2pt,
    thick,
    align=left,
    inner xsep=1.5mm,
    inner ysep=1mm,
    fill=black!3,
    minimum height=7mm,
    text width=24mm
  },
  proc/.style={iconbox, fill=orange!8},
  artf/.style={iconbox, fill=blue!7},
  outp/.style={iconbox, fill=green!10},
  role/.style={iconbox, fill=purple!8, text width=24mm, align=center},
  line/.style={-Latex, thick}
]
\node[role] (experts) {\faUsers\\ \textbf{Domain Experts}};
\node[proc, below=2mm of experts] (metamodel)
  {\faSitemap\ \textbf{Metamodels}\\ \emph{elements, relations, constraints}};
\node[proc, below=2mm of metamodel] (ontology)
  {\faProjectDiagram\ \textbf{Role-Metamodel Mapping}\\ \emph{semantic categories}};
\node[proc, below=2mm of ontology] (syntax)
  {\faShapes\ \textbf{Concrete Syntax}\\ \emph{notations and naming}};
\node[artf, anchor=north west] (specs)
  at ([xshift=9mm]experts.north east)
  {\faTasks\ \textbf{Problem Specs}\\ agreed with experts};
\node[outp, below=3mm of specs] (instances)
  {\faCubes\ \textbf{Reference Instances}\\ complete ground-truth models};
\node[outp, anchor=north west] (run)
  at ([xshift=9mm]specs.north east)
  {\faRobot\ \textbf{Completion Sessions}\\ run all configurations\\ capture traces};
\node[outp, below=3mm of run] (metrics)
  {\faChartBar\ \textbf{Metrics}\\ contribution, candidates, redundancy};
\begin{scope}[on background layer]
\node[lane, fit=(experts) (metamodel) (ontology) (syntax)] (L1) {};
\node[lane, fit=(specs) (instances)] (L2) {};
\node[lane, fit=(run) (metrics)] (L3) {};
\end{scope}
\node[title] at ([xshift=1.5mm,yshift=1.5mm]L1.north west)
  {Phase 1 -- Co-design};
\node[title] at ([xshift=1.5mm,yshift=1.5mm]L2.north west)
  {Phase 2 -- Ground Truth};
\node[title] at ([xshift=1.5mm,yshift=1.5mm]L3.north west)
  {Phase 3 -- Evaluation};
\draw[line] (experts) -- (metamodel);
\draw[line] (metamodel) -- (ontology);
\draw[line] (ontology) -- (syntax);
\draw[line] (metamodel.east) -- ++(4mm,0) |- (specs.west);
\draw[line] (ontology.east)  -- ++(4mm,0) |- (specs.west);
\draw[line] (syntax.east)    -- ++(4mm,0) |- (specs.west);
\draw[line] (specs) -- (instances);
\draw[line] (instances.east) -- ++(4mm,0) |- (run.west);
\draw[line] (run) -- (metrics);
\end{tikzpicture}
}
\caption{Offline evaluation protocol.}
\label{fig:offline-validation-workflow}
\end{figure}

Figure~\ref{fig:offline-validation-workflow} provides an overview of the offline evaluation protocol followed in this study. The protocol is organized into three phases: co-designing the DSL artifacts with domain experts, constructing expert-defined ground-truth models, and running the completion sessions under the different experimental configurations. The following subsections detail each part of this protocol.

\subsubsection{DSL Preparation and Ground Truth Construction}

We begin by collaborating with domain experts to define the 
metamodel for each DSL, capturing its structural elements, 
relationships, and constraints. We then define the role-metamodel mapping  by aligning metamodel elements with semantic categories, which supports context-sensitive prompt generation. Finally, we define the concrete syntax, including notations, graphical representations, and naming conventions, with continuous expert feedback throughout the process.

Once the DSLs are fully specified, we construct reference 
instances that serve as ground truth for evaluation. These 
complete models are built with domain experts based on agreed 
specifications and serve as the reference against which 
generated suggestions are evaluated.

\subsubsection{Domain-Specific Languages Under Study.}
  \label{sec:dsls}

\begin{table}[ht]
\footnotesize
\centering
\setlength{\tabcolsep}{4pt}
\renewcommand{\arraystretch}{1.02}
\begin{tabular}{ll}
\toprule
\textbf{Lang.} & \textbf{Scenarios} \\
\midrule
UML-CD   & Medical appointment;\\
         & Food delivery application \\
SCM      & Car manufacturing;\\
         & Aircraft maintenance routing \\
SysML-v2 & Flashlight assembly;\\
         & BatMobile system design \\
\bottomrule
\end{tabular}
\caption{Validation test cases.}
\label{tab:validation-cases}
\end{table}
To evaluate the applicability and effectiveness of our approach, we selected three domain-specific languages (DSLs) that differ in abstraction level, metamodel size, modeling paradigm, and industrial usage. \revadd{This selection was designed to exercise the main dimensions on which our approach relies: metamodel-based guidance, domain-semantic guidance, and applicability across heterogeneous DSLs.} 

\emph{UML Class Diagrams (UML-CD).} A widely used object-oriented structural language for software and data modeling. We use it to capture conceptual models for \emph{medical appointment management} and \emph{food delivery systems}. This DSL exemplifies schema-oriented modeling where class–attribute–association structures dominate.

\emph{Supply Chain Modeling (SCM).} A process-oriented DSL for representing logistics, production flows, and resource dependencies. We designed two industrial scenarios, \emph{car manufacturing} and \emph{aircraft maintenance routing}, to test the approach on operational process networks.

\emph{SysML-v2.} We use two illustrative scenarios, \emph{Flashlight assembly} and \emph{BatMobile system design}, to assess the approach on  component-based models. SysML~v2\footnote{We used the official SysML~v2 specification available at \url{https://github.com/Systems-Modeling/SysML-v2-Release/blob/master/doc/2a-OMG_Systems_Modeling_Language.pdf}}
 offers richer semantics and stricter typing, making it an ideal stress test for semantic-aware completion.

Table~\ref{tab:validation-cases} summarizes the selected DSLs and scenarios. Together, they cover diverse modeling styles (structural, behavioral, and systems-level) and support evaluating the assistant across several modeling languages. All specifications, metamodels, and ground-truth instances are provided in the replication package\footnote{\url{https://anonymous.4open.science/r/Evaluation_RAFiC}}.

\subsubsection{Simulating the Modeling Process}
\label{sec:simulation}

To mimic interactive modeling, we simulate an iterative completion process. Each evaluation starts from an \textit{initial partial model} derived from the ground truth by retaining only a minimal, syntactically valid subset of elements centered on the \emph{entrypoints} of the modeling language, as defined by the metamodel-based prior \(S_{\mathrm{metamodel}}\) (Section~\ref{sec:orchestration}). In practice, the model is initialized with zero or the top-1 instance of the highest-ranked entrypoint type, while all other ground-truth elements, relations, and characteristics are removed. At each iteration, a completion mechanism generates candidate suggestions from the current model state. 
These suggestions are evaluated against the ground truth and accepted or rejected based on \emph{semantic correspondence}, allowing equivalent or conceptually aligned elements. To limit subjectivity, decisions follow predefined guidelines established with domain experts, prioritizing semantic equivalence over exact naming; in ambiguous cases, suggestions are conservatively rejected.
The simulation stops when a fixed iteration limit is reached or no further valid suggestions can be identified.

\subsubsection{Evaluation Metrics}
\label{sec:evaluation-metrics}

\revadd{We evaluate each configuration by comparing the suggestions it produces with the expert-defined ground-truth model, allowing synonym-level matches. We define a \emph{ground-truth unit} as an atomic modeling contribution that can be recovered during completion. Depending on the DSL, such units may correspond to model elements, relations, or characteristics such as attributes, properties, or typed connections. Let \(G\) denote the set of all 
ground-truth units in the reference model.}

The \textbf{contribution rate} (\textbf{Ctr}) measures the 
percentage of the expected model recovered during the whole 
completion session. Let \(A\) denote the set of unique suggested units accepted as belonging to \(G\), either directly or through semantic correspondence:
\(Ctr = \frac{|A|}{|G|} \times 100\)

The \textbf{total unique number of generated candidates} 
(\textbf{Cand}) captures the overall amount of content proposed, independently of whether suggestions are accepted, rejected, or repeated. Let \(S_t\) denote the set of candidate units suggested at iteration \(t\) and \(T\) the total number of iterations: \( Cand = \sum_{t=1}^{T} |S_t| \)

\textbf{Redundancy} (\textbf{Red.}) measures how often a 
configuration repeatedly suggests the same modeling units. Let \(U_{\mathrm{cand}}\) denote the set of unique candidate units after merging identical or semantically equivalent suggestions:
\( Red = \frac{N_{\mathrm{cand}} - |U_{\mathrm{cand}}|}
{N_{\mathrm{cand}}} \times 100\)

In addition, we report the \textbf{cumulative number of accepted suggestions} over the interaction. Let \(a_i\) denote the number of new suggestions accepted at iteration \(i\): \( C_t = \sum_{i=1}^{t} a_i \).
This curve complements the aggregate metrics by showing how 
accepted contributions accumulate during the interaction, 
distinguishing configurations that make early progress from 
those that stagnate after a few iterations.

\subsubsection{Experimental Configurations}
\label{sec:configs}

We evaluate four configurations along two prompt-design axes: (i) \emph{structural injection}, which provides explicit structural guidance for the DSL, and (ii) \emph{semantic injection}, which provides meaning-level guidance via the role-metamodel mapping. For each axis, guidance may be provided \emph{iteratively}, \emph{as a whole} upfront, or not at all.
Different metrics  are computed separately for each configuration.
\begin{table}[!h]
\centering
\small
\setlength{\tabcolsep}{4pt}
\renewcommand{\arraystretch}{1.15}
\begin{tabular}{l l c c}
\hline
\textbf{Cfg} & \textbf{Configuration} & \textbf{Syntax injection} & \textbf{Semantic injection} \\
\hline
\textbf{\tool} & Tool-based completion & Iterative & Iterative \\
\textbf{\tool-Abl} & Tool-based ablation (no semantics) & Iterative & None \\
\textbf{C-Min} & Chat baseline (minimal prompt) & None & None \\
\textbf{C-Full} & Chat baseline (full prompt) & Whole (upfront) & Whole (upfront) \\
\hline
\end{tabular}
\caption{Experimental configurations along two prompt-design axes: syntax injection and semantic injection.}
\label{tab:configs-axes}
\end{table}

\paragraph{Tool-based completion.}
We evaluate our approach in two variants.  \textbf{\tool}(\emph{syntax+semantics, iterative}) injects structural constraints and semantic guidance derived from the annotation scheme at each iteration, using the current model state to tailor suggestions. \textbf{\tool-Abl.} (\emph{syntax-only, iterative}) keeps the same structural prompting but removes guidance from the semantic annotation scheme, serving as an ablation.

\paragraph{Chat-based completion.}
Following the chat-based protocol of \citet{camara2023assessment}, we define two baselines that generate suggestions through direct LLM prompting without orchestration. \textbf{C-Min} (\emph{no injections}) provides only the current partial model and the task instruction. \textbf{C-Full} (\emph{whole injections}) provides the current partial model together with the full metamodel and the semantic annotation scheme  injected upfront, rather than reconstructed iteratively. For reproducibility, the companion repository includes the exact prompts for C-Min/C-Full, configuration files, execution scripts, and raw logs\footnote{\url{https://github.com/meriembenchaaben/chat_based_completion}}. All four configurations are evaluated under identical conditions. For each DSL and scenario, they start from the same initial partial model, use the same entrypoint algorithm and acceptance policy, and are evaluated against the same ground truth. They also rely on the same locally deployed LLM, Llama 3.1-70B\footnote{https://huggingface.co/meta-llama/Llama-3.1-70B-Instruct}, with identical decoding parameters.

\section{Experimental results}

\label{sec:results}

\begin{figure*}[h]
\centering
  \includegraphics[width=0.65\textwidth]{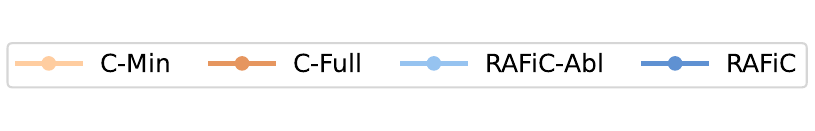}
  \vspace{-1mm}

  \begin{tabular}{c@{\hspace{0.02\textwidth}}|@{\hspace{0.02\textwidth}}c@{\hspace{0.02\textwidth}}|@{\hspace{0.02\textwidth}}c}
  \textbf{UML-CD} & \textbf{SCM} & \textbf{SysML-v2} \\[1mm]

  \begin{subfigure}[t]{0.28\textwidth}
    \centering
    \includegraphics[width=\linewidth]{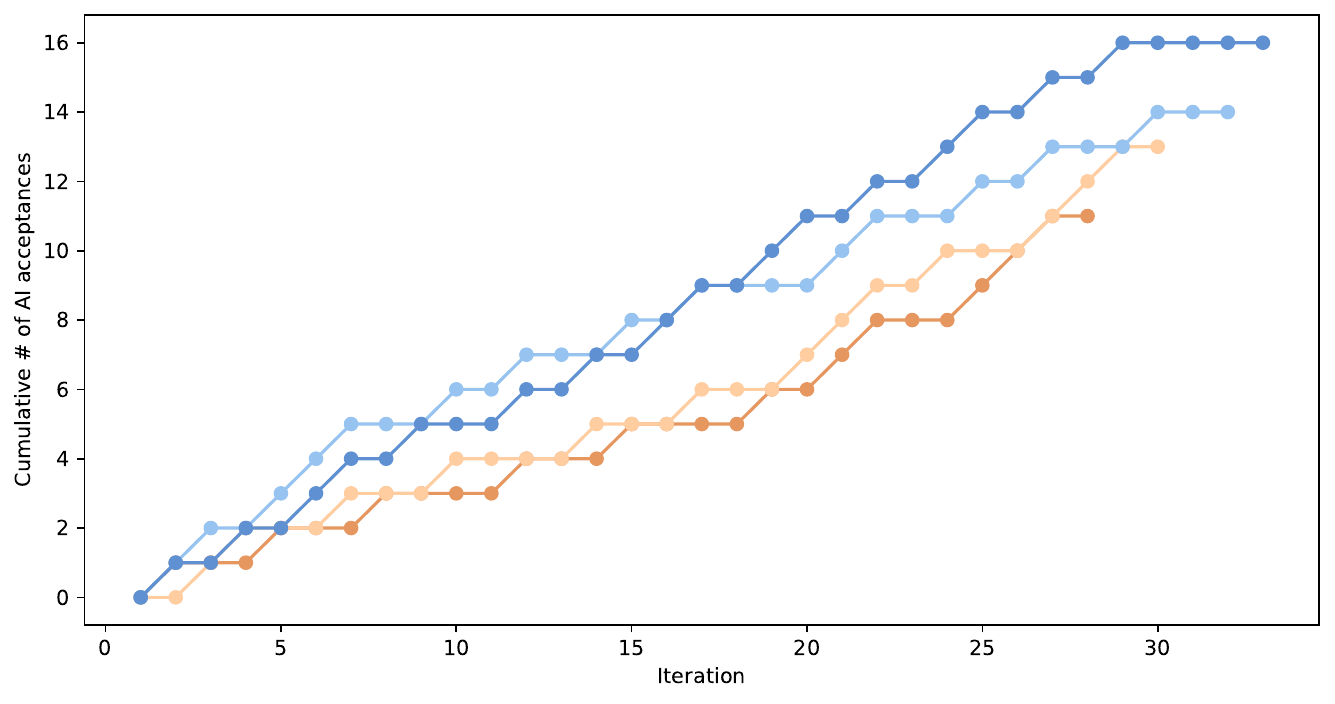}
    \caption{Food delivery}
    \label{fig:cd_delivery_cum_ai}
  \end{subfigure}
  &
  \begin{subfigure}[t]{0.28\textwidth}
    \centering
    \includegraphics[width=\linewidth]{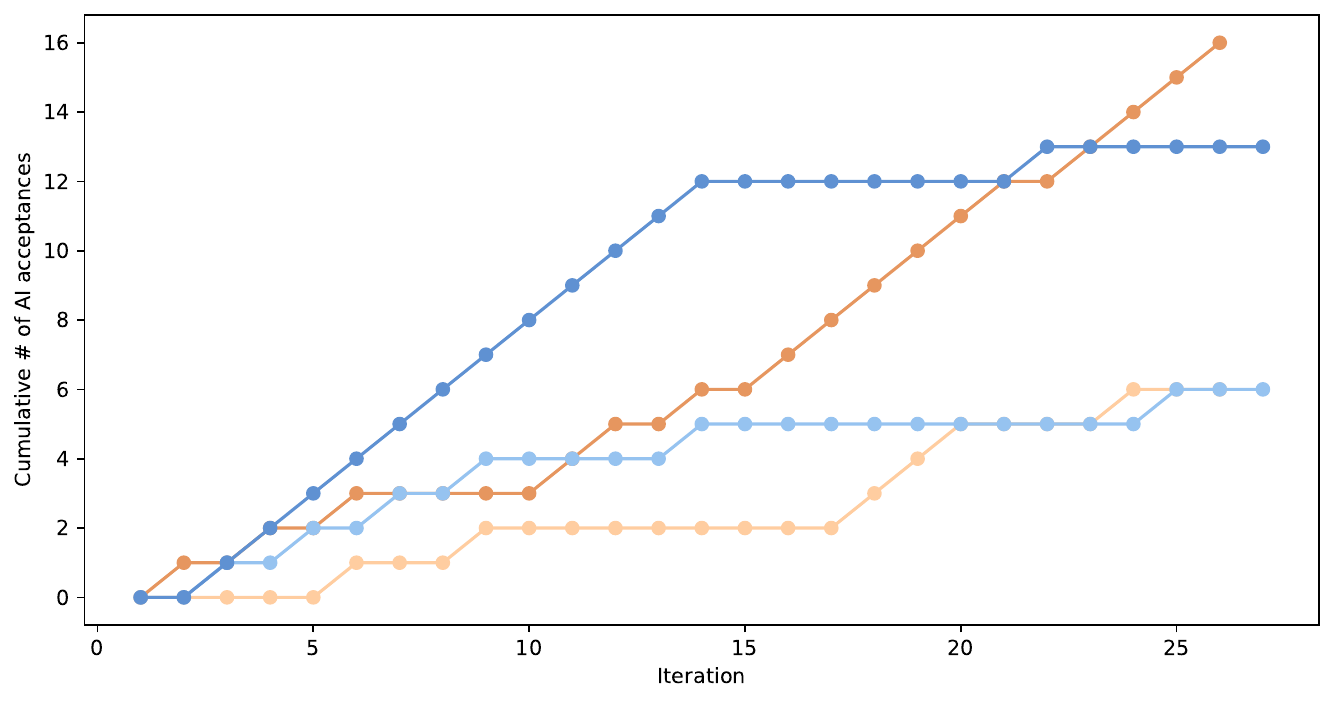}
    \caption{Car manufacturing}
    \label{fig:scm_car_cum_ai}
  \end{subfigure}
  &
  \begin{subfigure}[t]{0.28\textwidth}
    \centering
    \includegraphics[width=\linewidth]{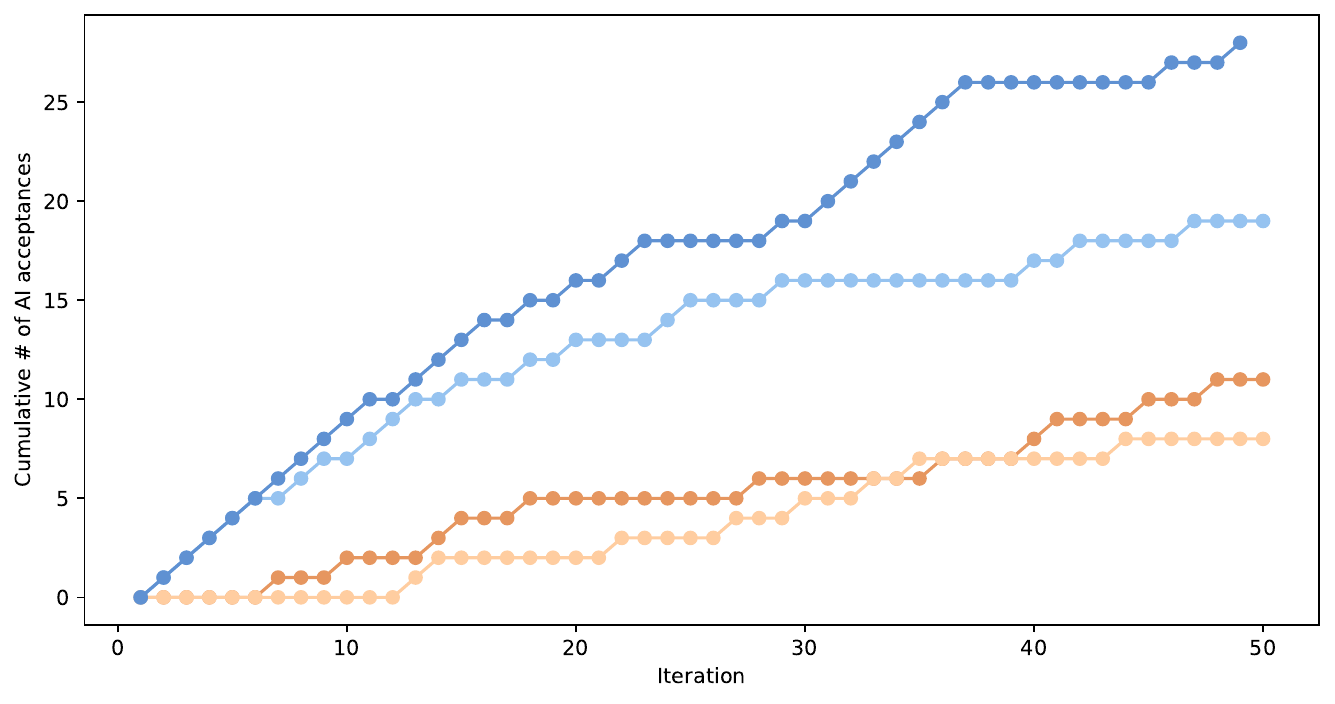}
    \caption{Flashlight}
    \label{fig:sysml_flashlight_cum_ai}
  \end{subfigure}
  \\[2mm]

  \begin{subfigure}[t]{0.28\textwidth}
    \centering
    \includegraphics[width=\linewidth]{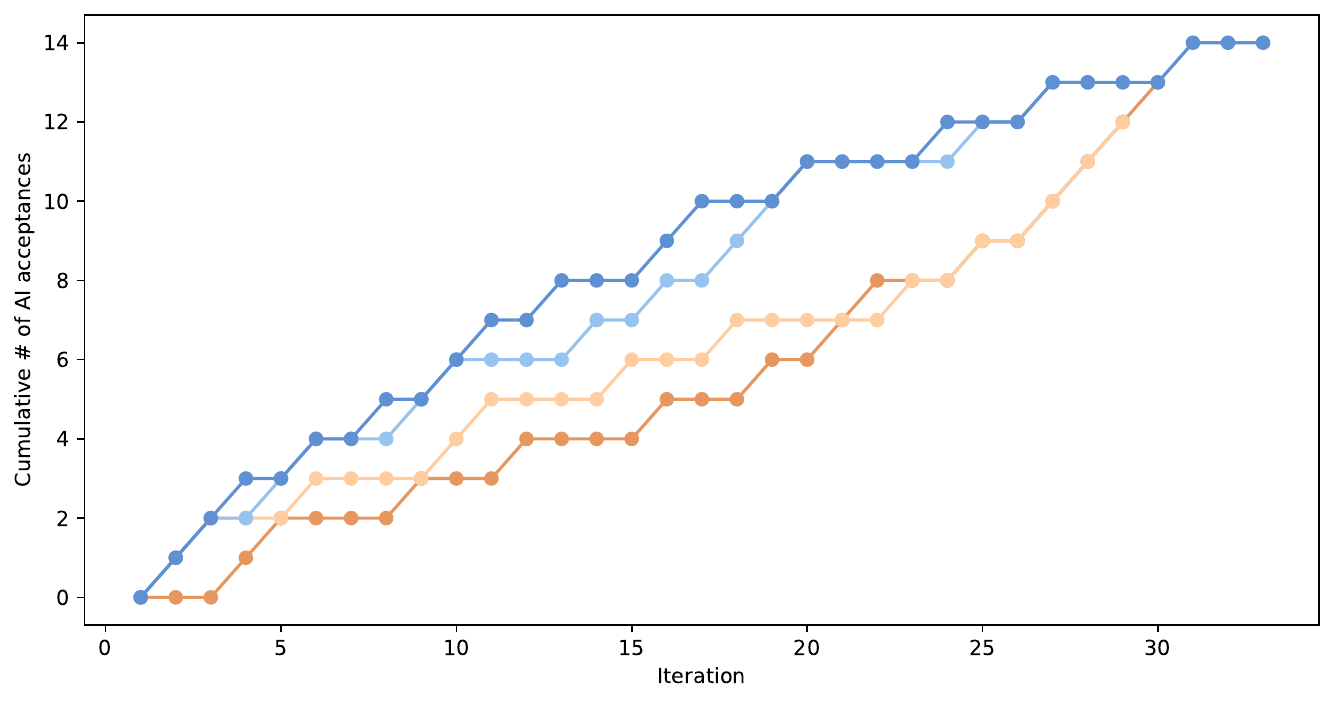}
    \caption{Medical appointment}
    \label{fig:cd_appointment_cum_ai}
  \end{subfigure}
  &
  \begin{subfigure}[t]{0.28\textwidth}
    \centering
    \includegraphics[width=\linewidth]{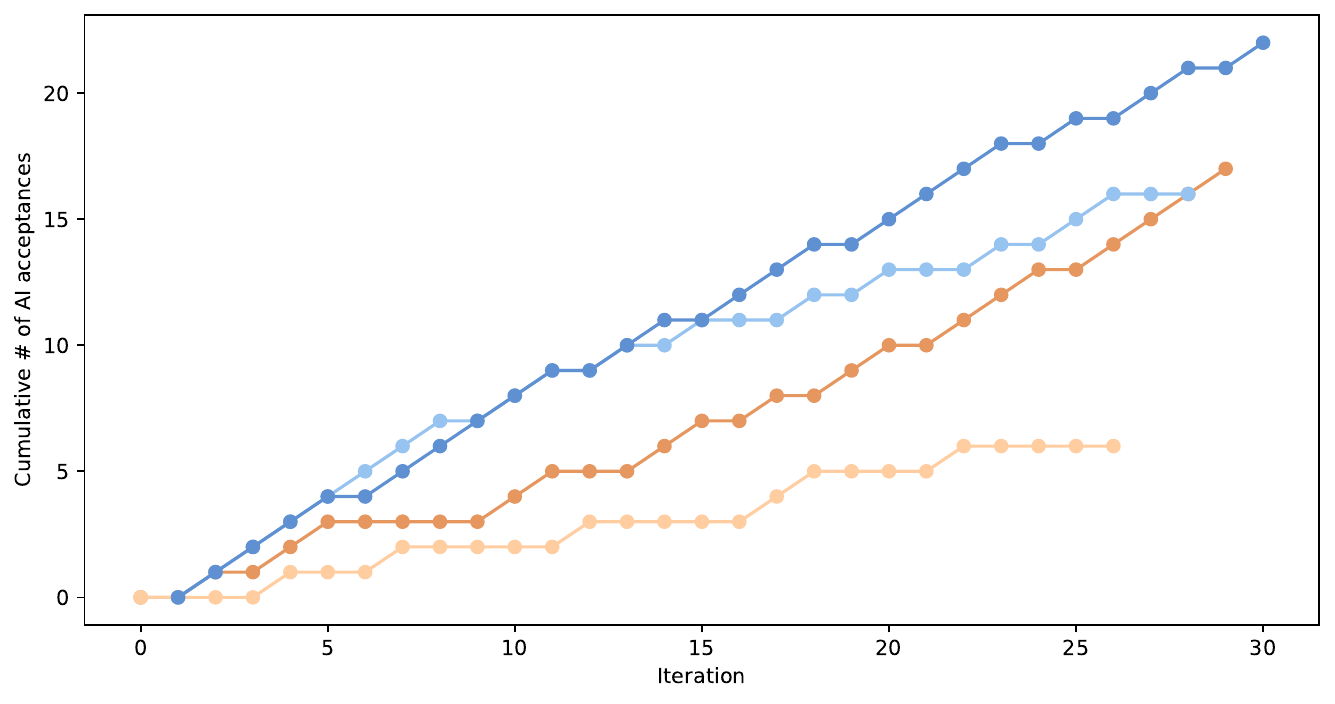}
    \caption{Aircraft maintenance}
    \label{fig:scm_aircraft_cum_ai}
  \end{subfigure}
  &
  \begin{subfigure}[t]{0.28\textwidth}
    \centering
    \includegraphics[width=\linewidth]{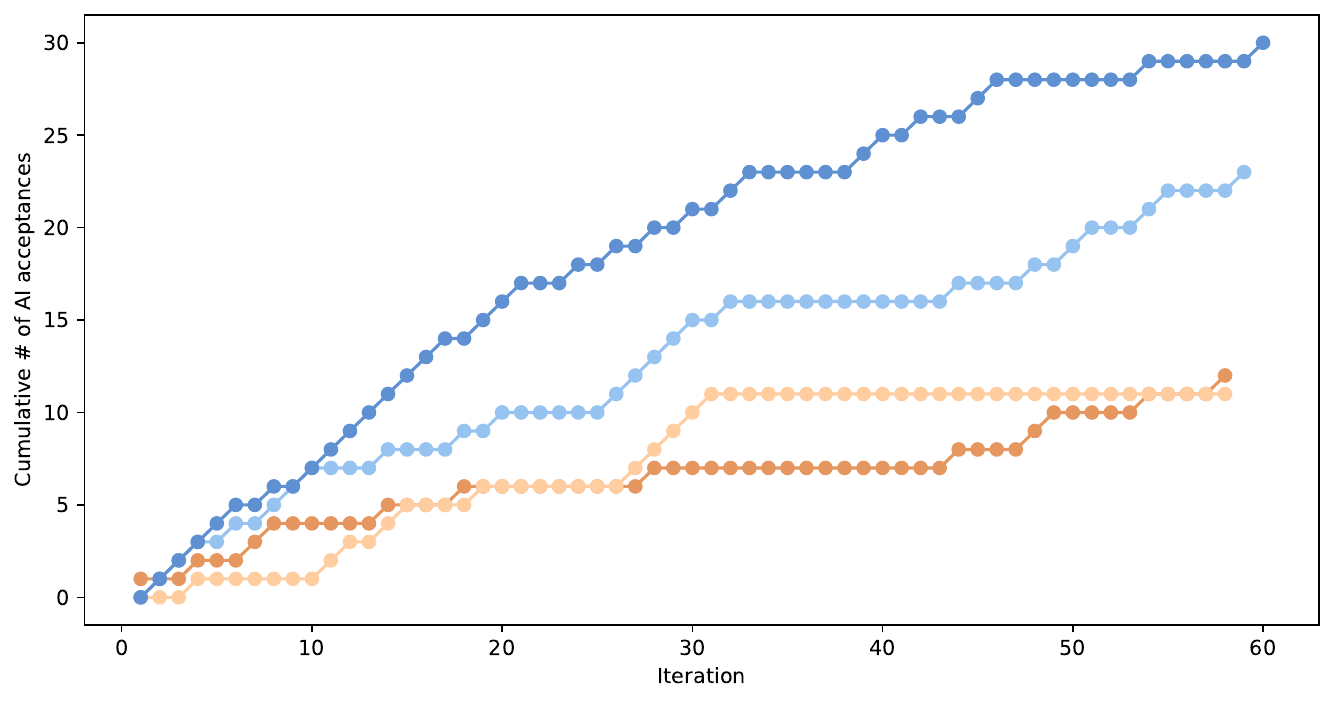}
    \caption{BatMobile}
    \label{fig:sysml_batmobile_cum_ai}
  \end{subfigure}
  \end{tabular}

  \caption{Cumulative AI-accepted suggestions across modeling progress for the DSLs and scenarios, grouped by DSL.}
  \label{fig:cum_ai_acceptances_all}
\end{figure*}

This section reports the empirical evaluation of \textbf{\tool} across the three DSLs introduced in Section~\ref{sec:dsls}. Table~\ref{tab:contr-cand-all-configs} summarizes the three metrics defined in Section~\ref{sec:evaluation-metrics} for all configurations: contribution rate (\textbf{Ctr.}), unique candidates (\textbf{Cand.}), and redundancy (\textbf{Red.}). Figure~\ref{fig:cum_ai_acceptances_all} complements these aggregate results by showing how accepted suggestions accumulate over the interaction, which helps interpret the progression patterns behind the final scores.

\definecolor{Tlight}{RGB}{150, 195, 240}  
\definecolor{Tdark}{RGB}{95, 145, 210}    

\definecolor{Clight}{RGB}{255, 205, 160}  
\definecolor{Cdark}{RGB}{230, 150, 95}    
\begin{table*}[h]
\centering
\small
\setlength{\tabcolsep}{1.8pt}
\renewcommand{\arraystretch}{1.0}
\begin{tabular}{l l
                >{\columncolor{Tdark}}r  >{\columncolor{Tdark}}r  >{\columncolor{Tdark}}r ||
                >{\columncolor{Clight}}r >{\columncolor{Clight}}r >{\columncolor{Clight}}r
                >{\columncolor{Cdark}}r  >{\columncolor{Cdark}}r  >{\columncolor{Cdark}}r
                >{\columncolor{Tlight}}r >{\columncolor{Tlight}}r >{\columncolor{Tlight}}r}
\hline
\textbf{DSL} & \textbf{Scen.} &
\multicolumn{6}{c}{\textbf{RQ1}} &
\multicolumn{3}{c}{\textbf{RQ2}} &
\multicolumn{3}{c}{\textbf{RQ3}} \\
\cline{3-14}
& &
\multicolumn{3}{c}{\textbf{\tool}} &
\multicolumn{3}{c}{\textbf{C-Min}} &
\multicolumn{3}{c}{\textbf{C-Full}} &
\multicolumn{3}{c}{\textbf{\tool-Abl}} \\
\cline{3-14}
& &
\shortstack{\textbf{Ctr.}\\\textbf{(\%)}} &
\shortstack{\textbf{Cand.}\\\textbf{(u.)}} &
\shortstack{\textbf{Red.}\\\textbf{(\%)}} &
\shortstack{\textbf{Ctr.}\\\textbf{(\%)}} &
\shortstack{\textbf{Cand.}\\\textbf{(u.)}} &
\shortstack{\textbf{Red.}\\\textbf{(\%)}} &
\shortstack{\textbf{Ctr.}\\\textbf{(\%)}} &
\shortstack{\textbf{Cand.}\\\textbf{(u.)}} &
\shortstack{\textbf{Red.}\\\textbf{(\%)}} &
\shortstack{\textbf{Ctr.}\\\textbf{(\%)}} &
\shortstack{\textbf{Cand.}\\\textbf{(u.)}} &
\shortstack{\textbf{Red.}\\\textbf{(\%)}} \\
\hline
\multicolumn{14}{l}{\textbf{UML-CD}}\\
& Food del.   & 39\%   & 122 & 42\% & 43.3\% & 20 & 65.5\% & 39\%   & 31 & 65.6\% & 34\% & 90  & 23\% \\
& Med. appt.  & 40\%   & 135 & 38\% & 44.8\% & 33 & 53.6\% & 43\%   & 27 & 56.5\% & 36\% & 100 & 25\% \\
\hline
\multicolumn{14}{l}{\textbf{SCM}}\\
& Car manuf.  & 48.1\% & 239 & 28\% & 23.1\% & 19 & 75.9\% & 56.7\% & 55 & 34.5\% & 23.3\% & 187 & 15\% \\
& Aircraft mt. & 74\%  & 294 & 54.1\% & 29\% & 70 & 42\% & 61\% & 75 & 31.8\% & 58\% & 239 & 48.82\% \\
\hline
\multicolumn{14}{l}{\textbf{SysML-v2}}\\
& Flashlight  & 57\%   & 155 & 37\% & 16\% & 45 & 44.4\% & 22\% & 52 & 25\% & 38\% & 128 & 13\% \\
& BatMobile   & 50\%   & 227 & 57\% & 19\% & 17 & 90.2\% & 20.7\% & 40 & 55\% & 39\% & 198 & 41\% \\
\hline
\end{tabular}%
\caption{Summary of results across DSLs and scenarios, grouped by configuration.}
\label{tab:contr-cand-all-configs}
\end{table*}

\subsection*{\textbf{RQ1: How much does \tooll improve over an interactive LLM-based completion baseline that does not incorporate DSL syntax or semantics?}}

RQ1 compares \textbf{\tool} with \textbf{C-Min}. In this  comparison, \textbf{\tool} improves contribution in 4 of the 6 scenarios, with the clearest gains in \textbf{SCM} and \textbf{SysML-v2}. For example, contribution rises from \textbf{29\%} to \textbf{74\%} in \emph{Aircraft maintenance} and from \textbf{16\%} to \textbf{57\%} in \emph{Flashlight}. By contrast, \textbf{C-Min} remains competitive in the two \textbf{UML-CD} scenarios, suggesting that simpler and more familiar modeling constructs can already be handled reasonably well by prompts that are minimally guided.

\revadd{
The candidate and redundancy metrics add nuance to this result. \textbf{\tool} generates more unique candidates in all scenarios and has lower redundancy in 5 of the 6 scenarios. This indicates that the improvement is not only due to producing more suggestions, but also to exploring the completion space more effectively. The main exception is \emph{Aircraft maintenance}, where higher contribution is accompanied by higher redundancy, suggesting a trade-off between semantic coverage and repeated exploration in more complex scenarios.
}

The cumulative curves confirm this, \textbf{\tooll}accumulates accepted 
suggestions earlier and more steadily, especially for Aircraft 
maintenance, Flashlight, and BatMobile scenarios.

\subsection*{\textbf{RQ2: How much does iterative, context-aware guidance in \tooll improve over an interactive LLM-based approach where DSL syntax and semantics are provided only once upfront?}}

RQ2 compares \textbf{\tool} with \textbf{C-Full}.  In this  comparison, providing DSL syntax and domain semantics only once upfront is not sufficient to consistently match iterative, context-aware guidance. \textbf{\tool} achieves higher contribution in 3 of the 6 scenarios, matches \textbf{C-Full} in 1, and is lower in 2. The clearest gains appear in the more complex settings, such as \emph{Flashlight}, where contribution rises from \textbf{22\%} to \textbf{57\%}, and \emph{Aircraft maintenance}, where it rises from \textbf{61\%} to \textbf{74\%}. By contrast, \textbf{C-Full} remains competitive in \emph{Car manufacturing} and \emph{Medical appointment}, while both configurations obtain the same contribution in \emph{Food delivery}. Overall, the benefit of iterative guidance is limited for \textbf{UML-CD}, mixed for \textbf{SCM}, and strongest for \textbf{SysML-v2}.

\revadd{
The candidate and redundancy metrics further qualify this result. \textbf{\tool} generates more unique candidates than \textbf{C-Full} in all scenarios, confirming that iterative guidance explores a broader completion space than a one-shot provision of syntax and semantics. However, this broader exploration is not always less redundant. \textbf{\tool} reduces redundancy in the two \textbf{UML-CD} scenarios and in \emph{Car manufacturing}, but shows higher redundancy in \emph{Aircraft maintenance}, \emph{Flashlight}, and \emph{BatMobile}. This suggests that, in more complex scenarios, iterative context-aware guidance can recover more useful completions, but may also revisit similar semantic areas as the model evolves.
}

\revadd{
 The cumulative curves show the same pattern, \textbf{\tooll}rises earlier and stays above \textbf{C-Full} for most of the interaction in the \textbf{SCM} and \textbf{SysML-v2} scenarios. Overall, iterative guidance matters most when useful completions depend on the evolving local context.
}

\subsection*{\textbf{RQ3: What is the contribution of the semantic role guidance within \textbf{\tool}, as assessed through an ablation study?}}

RQ3 compares \textbf{\tool} with \textbf{\tool-Abl}.  In this  comparison, the effect of semantic role guidance is clear: \textbf{\tool} achieves higher contribution, i.e., it recovers a larger proportion of the expert-defined ground-truth units, in all 6 scenarios. The gains are modest for \textbf{UML-CD}, but more pronounced for \textbf{SCM} and \textbf{SysML-v2}. For example, contribution increases from \textbf{58\%} to \textbf{74\%} in the \emph{Aircraft maintenance} scenario and from \textbf{38\%} to \textbf{57\%} in the \emph{Flashlight} scenario. This suggests that semantic guidance is especially useful when relevant completions depend on domain-specific concepts that are not captured by the DSL structure alone.

Semantic role  guidance also changes the explored completion space. In all 6 scenarios, \textbf{\tool} generates more unique candidates than \textbf{\tool-Abl}, indicating that semantic role  guidance expands the set of distinct completion alternatives considered by the system. However, this broader exploration is also accompanied by higher redundancy in all scenarios.
\revadd{
For example, redundancy rises from \textbf{13\%} to \textbf{37\%} in the \textbf{SysML-v2} \emph{Flashlight} scenario and from \textbf{23\%} to \textbf{42\%} in the \textbf{UML-CD} \emph{Food delivery} scenario. This suggests that semantic guidance helps identify more semantically relevant completions, but can also lead the system to revisit similar semantic areas or produce partially overlapping suggestions.
}

The cumulative curves support the same conclusion. In several scenarios, especially for \textbf{SCM} and \textbf{SysML-v2}, the \textbf{\tool} curve rises more strongly and reaches a higher final level than \textbf{\tool-Abl}. Taken together, these results indicate that semantic role guidance  substantially improves completion quality, with a trade-off in efficiency due to increased redundancy.

\section{Discussion}
\label{sec:discussion}

\revadd{This section interprets the main experimental findings and draws implications for the design of LLM-based modeling assistants.}

\paragraph{Iterative guidance improves contextual completion.}
\revadd{The comparison between \textbf{\tool} and the upfront-guidance baseline \textbf{C-Full} indicates that the benefit of iterative guidance is not about providing more information, but about providing the right information at the right time. When DSL syntax and semantic roles are injected upfront, the LLM receives all relevant knowledge but lacks a mechanism to focus on what matters now. As the model evolves, the initial context drifts from the current state, and the LLM cannot prioritize among the many valid but contextually irrelevant completions it could generate. }
\revadd{\textbf{\tooll}addresses this by reframing each generation step in the current model. The orchestrator recomputes scores, identifies the most pressing gap, and instantiates a prompt narrowed to that gap. This explains why the cumulative curves for \textbf{\tooll} rise earlier and more steadily in the SCM and SysML-v2 scenarios. This suggests that for the incremental modeling tasks studied here, adapting prompt content to the current model state can be more effective than providing all available knowledge upfront.}

\paragraph{Benefits vary across DSLs.}
\revadd{
The observed gains are not uniform across the three DSLs. This is consistent with recent findings showing that LLM knowledge of language constructs varies significantly depending on their representation in training data~\cite{delgado2026much}. For \textbf{UML-CD}, whose patterns are well-represented in pretraining corpora, even C-Min recovers substantial ground truth, as the LLM's already encoded knowledge acts as implicit guidance. For \textbf{SCM} and \textbf{SysML-v2}, this implicit knowledge is insufficient: their modeling patterns cannot be reliably inferred from general-purpose language understanding, and explicit metamodel and semantic role guidance  fills a genuine knowledge gap. The practical consequence is that the preparation effort for \textbf{\tooll} artifacts pays off most for specialized or proprietary DSLs, which are precisely the cases where modeling assistance is most needed.}

\paragraph{Role-metamodel mapping  improves semantic coverage.}
\revadd{ The ablation (\textbf{\tooll} vs. \textbf{\tool-Abl}) isolates the contribution of the semantic annotation scheme from the basic structure of the metamodel.  The role-metamodel mapping  provides information the metamodel alone does not convey: the modeling role of DSL elements.
This disambiguation operates through two mechanisms. First, it influences prompt construction: when a Ticket is mapped to Goal, the prompt frames it as an objective to be resolved, directing the LLM toward achievement-oriented completions. Second, it influences the scoring of candidate completions: the scoring function promotes completions that realize the semantic role patterns it predicts (such as Process/Role or Goal/Process), surfacing semantically meaningful next steps ahead of structurally valid but arbitrary alternatives.} \revadd{This distinction matters because metamodel conformance is necessary but not sufficient for useful completion. A conformant suggestion that introduces a random valid element is less useful than one that addresses a genuine semantic gap. }

\paragraph{Semantic coverage comes with redundancy.}
\revadd{
A recurring observation is that configurations with higher contribution rate also exhibit higher redundancy, particularly when semantic role guidance is active.  By promoting candidates that match semantic role patterns, the system concentrates exploration in semantically relevant regions, which increases the likelihood of accepted suggestions but also of revisiting similar modeling areas across iterations.
This redundancy is not a deficiency of the semantic role guidance  itself but rather reflects limited diversity control in the current prototype. The orchestration engine does not track which parts of the completion space have already been explored. Future versions should address this through semantic clustering of near-duplicate candidates, score penalties for suggestions similar to recently rejected ones, and adaptive focus shifting toward under-explored model areas.
}

\paragraph{Implications for LLM-based modeling assistance.}
\revadd{Overall, these results suggest that LLM-based model completion can be effective and applicable across the DSLs studied here when the generation process combines metamodel constraints with semantic-role-guided enrichment. The approach also demonstrates that useful modeling assistance is achievable with an open-source LLM, provided that structural and semantic guidance are explicitly integrated into the completion loop.}

\section{Threats to validity}
\label{sec:threats}

We discuss the threats to the validity of our study along three dimensions: construct, internal, and external validity. For each threat, we describe how we mitigated it.

\paragraph{Construct validity.}
Domain modeling is an open-world task, where there is rarely a single correct completion. A useful suggestion may differ from the reference model in naming, ordering, or level of detail while still expressing the same modeling intent, so strict syntactic matching would undercount valid suggestions. To mitigate this, suggestions were checked against the ground-truth model according to their intended meaning, the current modeling context, and the metamodel constraints, following predefined acceptance guidelines: a suggestion was counted as accepted only when it could be clearly associated with a missing ground-truth element.

\paragraph{Internal validity.}
First, one of the authors conducted the simulation and was familiar with both the tool and the target formalisms, which may have introduced bias in interaction decisions or in the interpretation of suggestions. To mitigate this, the human role was limited to verifying whether suggestions matched expected ground-truth elements, rather than judging their quality, and the same guidelines and procedure were applied across all configurations, DSLs, and scenarios. Future work could involve multiple independent checkers and report inter-rater agreement.
Second, each configuration was run once per DSL and scenario, so the values in Table~\ref{tab:contr-cand-all-configs} are single-run estimates with no variance or significance testing. This follows from the study design: each run is an interactive session with the tool in which every suggestion is checked against the ground truth, so a run cannot be cheaply repeated. To mitigate this, decoding was deterministic, with the same model and decoding parameters for all configurations. This limits but does not eliminate the threat, since minor non-determinism in the LLM serving stack is still possible; future work should evaluate more scenarios per DSL and report variance and statistical tests across them.

\paragraph{External validity.}
First, each DSL was evaluated on only two scenarios, which cannot capture the full diversity of modeling situations in practice and may limit generalizability. To mitigate this, we selected three distinct formalisms with increasing metamodel complexity (Section~\ref{sec:dsls}): UML-CD as a small and compact metamodel, SCM as a medium-sized formalism with richer process and resource constructs, and SysML-v2 as a larger and more complex formalism combining structural, behavioral, and parametric aspects, each paired with two scenarios from different domains. Second, results could be inflated if the LLM had memorized the studied scenarios during pretraining. To mitigate this, we chose languages and scenarios that are not among the most widely studied, while keeping the languages plausibly known to LLMs through public resources, as our goal is to assess whether such existing knowledge can be leveraged in a context-aware and structurally valid way. Finally, the semantic annotation scheme was derived by examining a limited set of modeling languages rather than through an exhaustive survey of the design space, so DSLs or domains outside those considered here may require roles or relations not currently in the scheme. To mitigate this, the approach is extensible by construction: the scheme can be incrementally refined, and role-metamodel mappings can be added progressively for new DSLs and domains without requiring changes to \tooll itself. Moreover, the three evaluated DSLs span distinct modeling paradigms, which reduces the risk that the scheme only fits a single notation.
\section{Related work}
\label{sec:related_work}

\revadd{
Early work on automatic model generation mainly relied on traditional NLP pipelines to transform textual requirements into UML models, especially class diagrams \cite{more2012generating,bashir2021modeling}. These approaches typically applied linguistic preprocessing steps such as sentence splitting, tokenization, part-of-speech tagging, syntactic parsing, and dependency analysis, followed by handcrafted rules or heuristics to identify candidate classes, attributes, operations, and relationships. While such techniques helped automate parts of the transition from requirements to models, they were often tailored to specific notations, input styles, and rule sets.} \revadd{ More recently, LLMs have been explored for software modeling tasks \cite{busch2023just, fill2023conceptual,chen2023automated}. These approaches differ from traditional NLP-based methods because they rely less on explicit linguistic rules and more on the generative capabilities of pretrained models. However, they still face related challenges: generated models must conform to the target modeling language, remain semantically aligned with the domain, and often require human or automated checks to detect invalid, incomplete, or redundant elements.
}

\revadd{Although most related work focuses on full model generation, our work targets interactive model completion from partial models. This distinction is important. In full generation, the model is usually produced from an initial textual specification and then inspected, corrected, or regenerated. In model completion, the model already exists and evolves through incremental suggestions that must remain consistent with the current modeling context.} Consequently, Table~\ref{tab:related-work-comparison} compares LLM-assisted domain modeling approaches along six dimensions that are directly related to the challenges addressed in this work.

We first report the \emph{target modeling language}, indicating whether an approach is tied to a specific notation, such as UML class diagrams,  goal models, or SysML diagrams, or whether it can be applied to several languages when a metamodel is available, as in Ecore-based DSLs~\citep{Pan25}.

The \emph{LLM} column indicates whether an approach relies on proprietary cloud-based models as in many ChatGPT- or GPT-based studies~\citep{chen2023automated,camara2023assessment}, or considers open-source models  (OSS)  that can be executed locally.

The \emph{generation strategy} indicates whether models are produced in a single LLM invocation and subsequently inspected or accepted by a human, or through an iterative process involving multiple LLM interactions for refinement, with human and/or automated feedback.

\emph{Tool integration} captures whether the LLM support is embedded within a modeling environment or used externally, for example through a chat interface, an offline pipeline, or a separate textual notation such as PlantUML.

\revadd{Finally, \emph{syntax injection} indicates how an approach helps ensure that the generated or accepted modeling elements are valid with respect to the target modeling language. This may rely on explicit metamodel information, tool-supported checks, language rules, or additional constraints such as OCL. 
In contrast, \emph{semantic injection} indicates how domain meaning is provided to guide the relevance of these elements. This may be done through natural-language requirements, illustrative examples, scenario descriptions, or, in our case, semantic-role prompting that relates domain concepts to possible modeling elements.}

\begin{table}[t]
\centering
\scriptsize
\renewcommand{\arraystretch}{1.08}
\setlength{\tabcolsep}{2.0pt}
\rowcolors{2}{gray!6}{white}

\newcolumntype{Y}{>{\raggedright\arraybackslash}X}

\begin{tabularx}{\columnwidth}{@{}lcYYYYYY@{}}
\hline
\textbf{Work} &
\textbf{Year} &
\textbf{Target language} &
\textbf{LLM} &
\textbf{Generation} &
\textbf{Tool integration} &
\textbf{Syntax inj.} &
\textbf{Semantic inj.} \\
\hline

\citet{chen2023automated} &
2023 &
UML Class Diagrams &
GPT-3.5, GPT-4 (Comm) &
Non-iterative, acceptance-loop &
No editor integration; offline pipeline &
Implicit format constraints &
NL descriptions and examples \\

\textbf{\citet{camara2023assessment}} &
2023 &
UML Class Diagrams &
ChatGPT (Comm) &
\textbf{Iterative, HITL} &
No editor integration; PlantUML/USE outputs &
OCL constraints &
None \\

\citet{chen2023use} &
2023 &
Goal Models &
GPT-4 (Comm) &
Iterative, HITL &
No editor integration; interactive ChatGPT prompting &
None &
Natural-language descriptions \\

\citet{chaaben2024Utility} &
2024 &
UML Class Diagrams &
GPT-3 (Comm) &
Iterative, HITL &
MAGDA tool; Eclipse/Sirius-based environment &
Implicit constraints &
Few-shot examples \\

\citet{Ferrari24} &
2024 &
UML Sequence Diagrams &
ChatGPT (GPT-3.5/4, Comm) &
Non-iterative, acceptance-loop &
No editor integration; PlantUML output in PlantText &
None &
Natural-language requirements \\

\citet{Apvrille2024SystemAA} &
2024 &
SysML Diagrams &
ChatGPT (GPT-3.5, Comm) &
Iterative, automated feedback &
TTool-AI modeling and verification environment &
SysML rules and tool checks &
Natural-language specifications \\

\citet{Pan25} &
2025 &
Ecore-based DSLs &
GPT-4o, o1-preview (Comm), Llama~3.1 (8B, 70B) (OSS) &
Non-iterative, automated &
Prototype integration with EMF toolchain &
Explicit metamodel &
Scenario descriptions in NL \\

\rowcolor{blue!15}
\textbf{\tool} &
\textbf{This work} &
Any language with a metamodel &
Llama 3.1-70B (OSS) &
Iterative, HITL &
Prototype Sirius Web integration &
Explicit metamodel &
Semantic-role-guided prompting \\
\hline
\end{tabularx}

\vspace{2pt}
\tiny\emph{Comm. = commercial; OSS = open-source; HITL = human-in-the-loop.}
\caption{Comparison of LLM-assisted domain modeling approaches.}
\label{tab:related-work-comparison}
\end{table}

\revadd{
Among these works, \citet{camara2023assessment} is particularly relevant to our experimental design. Their work focuses on UML class diagrams and studies LLM-based model completion in an interactive, human-in-the-loop setting, where the LLM proposes modeling elements and the user progressively validates or rejects them. They identify significant limitations in semantic accuracy, particularly for larger diagrams, noting serious errors for models larger than 10--12 classes. We use this interaction style as inspiration for our chat-based baseline configurations described in Section~\ref{sec:configs}.
}

Our work differs from prior approaches by targeting language-agnostic, iterative model completion using an open-source LLM. Rather than focusing on a specific DSL or diagram type, \tooll supports any modeling language for which a metamodel is provided. The LLM is guided by both explicit metamodel information and semantic role guidance, enabling suggestions that are syntactically valid and semantically aligned across heterogeneous languages.  \revadd{Furthermore, unlike most related approaches that generate models in a text-based format (e.g., PlantUML), \tooll produces suggestions as actionable editing operations within a graphical model editor, ensuring that accepted recommendations are directly integrated into the model under construction without requiring manual transcription or format conversion.}

Beyond the approaches compared above, other aspects of model completion have also been studied. Logic-based encodings of metamodel knowledge and constraints have been used to support completion and correctness, enabling automated reasoning about properties of class models~\cite{berardi2005reasoning}. More recently, \citet{welter2026nextfocus} propose \emph{NextFocus}, a predictor that recommends \emph{where} to continue editing by ranking candidate focus nodes for \emph{multi-location} model completion from evolution history. Our work is complementary to NextFocus: rather than predicting where the modeler should focus next, it addresses \emph{what} to add, generating domain-relevant elements and connections and assessing their semantic adequacy via metamodel knowledge and role-metamodel mappings.

\section{Conclusion}
\label{sec:conclusion}

In this paper, we presented \tool, an LLM-based approach for iterative 
domain model completion that generalizes across DSLs. The approach combines 
metamodel-based structural constraints with semantic-role-guided
enrichment, and uses the evolving partial model as context to produce 
incremental, context-aware suggestions during model construction. 
We assess the approach through a comparative ablation study on three DSLs (UML-CD, SCM, and SysML-v2), each with two scenarios. The evaluation yielded three main findings. First, within the three DSLs and six scenarios evaluated, iterative guidance, which adapts prompts to the current model state, outperforms minimal prompting in 4 of the 6 scenarios and matches or outperforms one-shot upfront provision of DSL knowledge in 4 of the 6 scenarios, with the largest gains observed for DSLs less represented in LLM training data. Second, semantic role guidance improves contribution rate across all scenarios by providing modeling role information that the metamodel alone does not convey. Third, these results are achieved with an open-source LLM (LLaMA 3.1-70B), showing that clear structural and semantic guidance can help open-source models perform well even when they have less prior knowledge than commercial models.

These findings support a broader principle for LLM-based modeling tools: combining generative capabilities with explicit, modeling-specific knowledge produces more relevant suggestions than relying on the LLM's implicit knowledge alone. The approach is not tied to a specific language or notation, and can be applied to any DSL for which a metamodel and role-metamodel mapping are provided.

\revadd{
Future work will build on these results in three directions. First, evaluating \tooll on additional DSLs and with larger or commercial LLMs would test whether the observed patterns hold across a broader range of modeling languages and model 
capabilities. Second, user studies with modelers of varying expertise would assess how the coverage gains observed in our experiments translate into perceived usefulness and productivity during real modeling sessions. Third, providing semi-automated support for the role-metamodel mapping, for example by using the LLM itself to propose initial alignments, would reduce 
the preparation effort and facilitate adoption of \tooll for new DSLs.}

\bibliographystyle{ACM-Reference-Format}
\bibliography{references}

\end{document}